\documentclass[dvipdfmx]{pasj01}
\SetRunningHead{Okumura et al.}{Subaru FMOS Galaxy Redshift Survey (FastSound). IV. Redshift space distortions}

\usepackage{times}

\newcommand{\himpc}{{\hbox {$~h^{-1}$}{\rm ~Mpc}}}
\newcommand{\hmpci}{{\hbox {$~h{\rm ~Mpc}^{-1}$}}}

\newcommand{\vecr}{{\bf r}}
\newcommand{\vecx}{{\bf x}}
\newcommand{\veck}{{\bf k}}
\newcommand{\nn}{\nonumber}
\newcommand{\be}{\begin{equation}}
\newcommand{\ee}{\end{equation}}
\newcommand{\bey}{\begin{eqnarray}}
\newcommand{\eey}{\end{eqnarray}}
\newcommand{\hikage}{Hikage et al. in preparation}
\newcommand{\hikaget}{Hikage et al. (in preparation) }
\newcommand{\hikagep}{(Hikage et al. in preparation)}

\begin{document}

\title{The Subaru FMOS galaxy redshift survey (FastSound). IV.  
New constraint on gravity theory from redshift space distortions at $z\sim 1.4$}
\author{
Teppei \textsc{Okumura},\altaffilmark{1,*} 
Chiaki \textsc{Hikage},\altaffilmark{1}
Tomonori \textsc{Totani},\altaffilmark{2}\\
Motonari \textsc{Tonegawa},\altaffilmark{2}
Hiroyuki \textsc{Okada},\altaffilmark{2}
Karl \textsc{Glazebrook},\altaffilmark{3}\\
Chris \textsc{Blake},\altaffilmark{3}
Pedro G. \textsc{Ferreira},\altaffilmark{4}
Surhud \textsc{More},\altaffilmark{1}
Atsushi \textsc{Taruya},\altaffilmark{1,5} 
Shinji \textsc{Tsujikawa},\altaffilmark{6} 
Masayuki \textsc{Akiyama},\altaffilmark{7}
Gavin \textsc{Dalton},\altaffilmark{8,9}\\
Tomotsugu \textsc{Goto},\altaffilmark{10} 
Takashi \textsc{Ishikawa},\altaffilmark{11} 
Fumihide \textsc{Iwamuro},\altaffilmark{11} \\
Takahiko \textsc{Matsubara},\altaffilmark{12,13}
Takahiro \textsc{Nishimichi},\altaffilmark{1,14}
Kouji \textsc{Ohta},\altaffilmark{11}\\
Ikkoh \textsc{Shimizu},\altaffilmark{15}
Ryuichi \textsc{Takahashi},\altaffilmark{16} 
Naruhisa \textsc{Takato},\altaffilmark{17}\\
Naoyuki \textsc{Tamura},\altaffilmark{1}
Kiyoto \textsc{Yabe},\altaffilmark{1}
and Naoki \textsc{Yoshida}\altaffilmark{1,14,18}
}

\altaffiltext{1}{Kavli Institute for the Physics and Mathematics of the Universe (WPI), The University of Tokyo Institutes for Advanced Study, The University of Tokyo, 5-1-5 Kashiwanoha, Kashiwa 277-8583, Japan}
\altaffiltext{2}{Department of Astronomy, School of Science, The University of Tokyo, 7-3-1 Hongo, Bunkyo-ku, Tokyo 113-0033, Japan}
\altaffiltext{3}{Centre for Astrophysics \& Supercomputing, Swinburne University of Technology, P.O. Box 218, Hawthorn, VIC 3122, Australia}
\altaffiltext{4}{Department of Physics, University of Oxford, Denys Wilkinson Building, Keble Road, Oxford, OX13RH, United Kingdom}
\altaffiltext{5}{Yukawa Institute for Theoretical Physics, Kyoto University, Kyoto 606-8502, Japan}
\altaffiltext{6}{Department of Physics, Faculty of Science, Tokyo University of Science, 1-3 Kagurazaka, Shinjuku-ku, Tokyo 162-8601, Japan}
\altaffiltext{7}{Astronomical Institute, Faculty of Science, Tohoku University, 6-3 Aramaki, Aoba-ku, Sendai, Miyagi 980-8578, Japan}
\altaffiltext{8}{Astrophysics, Department of Physics, Denys Wilkinson Building, Keble Road, Oxford OX1 3RH, U.K.}
\altaffiltext{9}{RALSpace, STFC Rutherford Appleton Laboratory, HSIC, Oxford OX11 0QX, UK}
\altaffiltext{10}{Institute of Astronomy, National Tsing Hua University, No. 101, Section 2, Kuang-Fu Road, Hsinchu, Taiwan 30013}
\altaffiltext{11}{Department of Astronomy, Kyoto University, Kitashirakawa-Oiwake-cho, Sakyo-ku, Kyoto 606-8502, Japan}
\altaffiltext{12}{Department of Physics, Nagoya University, Furo-cho, Chikusa-ku, Nagoya 464-8602, Japan}
\altaffiltext{13}{Kobayashi-Maskawa Institute for the Origin of Particles and the Universe (KMI), Nagoya University, Furo-cho, Chikusa-ku, Nagoya 464-8602, Japan}
\altaffiltext{14}{CREST, JST, 4-1-8 Honcho, Kawaguchi, Saitama, 332-0012, Japan}
\altaffiltext{15}{Department of Earth \& Space Science, Graduate School of Science, Osaka University, 1-1 Machikaneyama, Toyonaka, Osaka 560-0043, Japan}
\altaffiltext{16}{Faculty of Science and Technology, Hirosaki University, 3 Bunkyo-cho, Hirosaki, Aomori 036-8561, Japan}
\altaffiltext{17}{Subaru Telescope, National Astronomical Observatory of Japan, 650 North A'ohoku Pl., Hilo, HI 96720, USA}
\altaffiltext{18}{Department of Physics, School of Science, The University of Tokyo, 7-3-1 Hongo, Bunkyo, Tokyo 113-0033, Japan}
\email{teppei.okumura@ipmu.jp}


\maketitle
 
\begin{abstract}
We measure the redshift-space correlation function from a spectroscopic sample of $2783 $ emission line galaxies from the FastSound survey. 
The survey, which uses the Subaru Telescope and covers the redshift ranges of $1.19<z<1.55$, is the first cosmological study at such high redshifts. 
We detect clear anisotropy due to redshift-space distortions (RSD) both in the correlation function as a function of separations parallel and perpendicular to the line of sight and its quadrupole moment.
RSD  has been extensively used to test general relativity on cosmological scales at $z<1$. 
Adopting a $\Lambda$CDM cosmology with the fixed expansion history and no velocity dispersion $\sigma_{\rm v}=0$, and using the RSD measurements on scales above $8\himpc$, we obtain the first constraint on the growth rate at the redshift, $f(z)\sigma_8(z)=0.482\pm 0.116 $ at $z\sim 1.4$ after marginalizing over the galaxy bias parameter $b(z)\sigma_8(z)$. This corresponds to $4.2\sigma$ detection of RSD. Our constraint is consistent with the prediction of general relativity $f\sigma_8\sim 0.392$ within the $1-\sigma$ confidence level. 
When we allow $\sigma_{\rm v}$ to vary and marginalize it over, the growth rate constraint becomes $f\sigma_8=0.494^{+0.126}_{-0.120}$.
We also demonstrate that by combining with the low-$z$ constraints on $f\sigma_8$, high-$z$ galaxy surveys like the FastSound can be useful to distinguish modified gravity models without relying on CMB anisotropy experiments.
\end{abstract}

\section{Introduction}

Probing the nature of dark energy, which drives the acceleration of the cosmic expansion, is one of the most important issues in cosmology (see \cite{Weinberg:2013} for the latest review).
The simplest explanation for the acceleration is to posit the existence of a cosmological constant. An alternative possibility is to assume the presence of an extra degree of freedom in the form of quintessence or its variants (e.g., \cite{Copeland:2006}) or to  modify Einstein's general relativity\footnote{This paper is completed and submitted on November 25, 2015, the 100th anniversary of Einstein's theory of general relativity.} on cosmological scales (e.g., \cite{Clifton:2012}). 
While geometric probes such as Type Ia supernovae \citep{Riess:1998,Perlmutter:1999} or baryon acoustic oscillations \citep{Eisenstein:2005} are useful to constrain the nature of dark energy, it is hard to break its degeneracy with modified gravity. On the other hand, density perturbations grow at different rates in dark energy and modified gravity.
Further observations of the growth rate at cosmological scales are thus required to distinguish between dark energy and modified gravity.

Among the various cosmological observations, large-scale structure surveys are considered to be extremely  powerful tools. Through the measurements of spectroscopic redshifts, galaxy redshift surveys provide 3-dimensional clustering information using galaxies as test particles. Moreover, since the redshift of a galaxy is a combination of expansion velocity and the galaxy's own motion, the radial positions of the galaxies inferred from the redshift are misplaced along the line of sight in what is known as redshift-space distortion (RSD) \citep{Sargent:1977,Kaiser:1987,Hamilton:1992,Cole:1994,Hamilton:1998}. By statistically analyzing the anisotropy in the galaxy clustering caused by the radial components of galaxy peculiar velocities, one can obtain additional information. 
RSD has been analyzed for various redshift surveys (e.g.,\cite{Davis:1983,Fisher:1994,Peacock:2001,Zehavi:2002,Hawkins:2003,Percival:2004,Tegmark:2004a,da-Angela:2005,Tegmark:2006,Ross:2007,Okumura:2008}).
These studies constrained the so-called linear redshift distortion parameter $\beta\equiv f/b$, where $f=d\ln D/d \ln a$ is the logarithmic derivative of the linear growth rate $D$ and $b$ is the linear bias parameter. 

\citet{Guzzo:2008} focused on the fact that cosmological models in different gravity theories that predict a similar expansion rate can have different values of $f$ via a change to the Poisson equation \citep{Wang:1998,Linder:2005} and demonstrated that it is potentially possible to distinguish different theories of gravity using RSD. 
Following their finding, constraints on $f$ or the normalized growth rate $f\sigma_8$ have been obtained for various surveys to examine any possible deviation of the gravity's law from general relativity at various redshifts; Two-Degree Field Galaxy Redshift Survey (2dFGRS) \citep{Song:2009a}, the 6dFGS \citep{Beutler:2012}, the Sloan Digital Sky Survey (SDSS) main galaxy sample \citep{Howlett:2015}, the SDSS luminous red galaxy (LRG) sample \citep{Yamamoto:2008, Cabre:2009,Samushia:2012}, the WiggleZ survey \citep{Blake:2011,Contreras:2013},  
the SDSS-III Baryon Oscillation Spectroscopic Survey (BOSS) \citep{Reid:2012,Tojeiro:2012}, the SDSS-III LOWZ sample \citep{Chuang:2013} and the VIMOS Public Extragalactic Redshift Survey (VIPERS) \citep{de-la-Torre:2013}, and 
currently good agreement with general relativity has been obtained.

Given that the presence of dark energy or modified gravity changes the gravitational assembly history of matter in the universe, it is important to obtain the constraints on the growth rate over a wide range of redshifts, that is, as a function of redshift $f(z)$. However,
reliable measurements of RSD are currently limited to the redshift $z<1$ 
(see e.g., figure 19 of \cite{de-la-Torre:2013})
although higher-redshift clustering can be probed using quasar or Lyman-$\alpha$ forest samples. 
It becomes very difficult to observe the galaxy distribution at redshift higher than 1 up to around $z\sim 2$ with optical surveys \citep{Steidel:2004}, thus it is hard to perform a test of gravity theories at such redshifts. 
Nevertheless, as emphasized in \citet{Tonegawa:2015}, at such higher redshift nonlinearities on the physical scales of interest are smaller than they are today.
This enables one to measure $f\sigma_8$ in a relatively unbiased fashion. Moreover, since $f$ approaches 1 at higher redshifts, RSD is directly sensitive to $\sigma_8$ and provides a baseline for lower redshift measurements. 
See \citet{Bielby:2013} for a constraint on the growth rate $f$ at $z\sim 3$ although the error on $f$ is $\sim 50\%$ 
($2\sigma$ detection of RSD). 

This is a part of a series of papers for the analysis of the FastSound survey. The FastSound survey is a near-infrared galaxy survey and measures redshifts of the galaxies using H$\alpha$ emission lines obtained with 
the Fiber Multi-Object Spectrograph (FMOS), a spectrograph equipped at Subaru Telescope. 
Paper I describes the overview of the FastSound survey \citep{Tonegawa:2015}. Paper II presents properties of emission line galaxies and explains our catalogs of the FastSound \citep{Okada:2015}. Paper III studies the mass-metallicity relation of the H$\alpha$ emission line galaxies \citep{Yabe:2015}.
This paper is the fourth paper of the series and we present the clustering analysis at redshift $1.2<z<1.5$ using the star-forming galaxy sample obtained from the FastSound survey. 
We measure the redshift-space two-point correlation function, derive a constraint on $f\sigma_8$
and perform a consistency test of gravity theory based on a $\Lambda$CDM model with general relativity. 

The structure of this paper is as follows. 
Section \ref{sec:data} describes the galaxy sample of the FastSound survey. 
We then measure the correlation function in redshift space and its covariance matrix in section \ref{sec:measurement}. 
In section \ref{sec:basics}, we review theoretical model predictions of the correlation function in redshift space, used in our analysis, and test them against mock catalogs based on $N$-body simulations. 
Constraints on the growth rate $f\sigma_8$ are obtained and deviation from general relativity is tested in section \ref{sec:analysis}. 
In section \ref{sec:fsigma8_z}, we present $f\sigma_8$ as a function of redshift and demonstrate what kind of insights we can obtain for gravity theory models by adding the observation of the FastSound survey.
We conclude in section \ref{sec:conclusion}. 
Several systematic effects on the growth rate constraints, such as constructions of the selection functions, are tested in section appendix \ref{sec:systematics}.


\section{Data}\label{sec:data}

The comprehensive overview of the FastSound
survey\footnote{http://www.kusastro.kyoto-u.ac.jp/Fastsound/index.html}
is presented in Paper I, and the properties of the emission line
galaxies obtained from the FastSound galaxy catalog are detailed in Paper
II (see also \cite{Sumiyoshi:2009,Tonegawa:2014,Tonegawa:2015a}).  In
this section we describe a brief overview of the FastSound survey, the
galaxy sample used for our cosmological analysis and the selection
functions used to create the random galaxy catalog without clustering.

\subsection{The FastSound galaxy data sample}\label{sec:fastsound}

The FastSound survey uses the near-infrared Fiber Multi-Object
Spectrograph (FMOS; \cite{Kimura:2010}) mounted on the Subaru
Telescope, which has about 400 fibers in a circular field of view
(FOV) of a 30 arcmin diameter.  The light from targets is sent to two
spectrographs, IRS1 and IRS2, each of which produces about 200 spectra.
The aim of the survey is to construct 3D maps of star forming galaxies
at $z = 1.19$--1.55, by detecting their redshifted H$\alpha$ emission
lines.  Targets for FMOS spectroscopy were selected from the four
fields (W1-W4) of the Canada-France-Hawaii Telescope Legacy Survey
(CFHTLS)\footnote{http://www.cfht.hawaii.edu}, which covers $\sim 170
{\rm \ deg}^2$ in total with five photometric band filters, $u$, $g$,
$r$, $i$ and $z$. (\cite{Ilbert:2006,Coupon:2009}; Goranova et al. 
2009\footnote{Y. Goranova et al. 2009, The CFHTLS T0006 Release (http://terapix.iap.fr/cplt/T0006-doc.pdf)};
\cite{Gwyn:2012}; Paper I ).
The FastSound covers 1.81, 6.62, 9.10, and 3.10 deg$^2$ of
CFHTLS W1--4, respectively, in total 20.61 deg$^2$, by 10, 39, 54, and
18 FMOS FOVs. Emission lines were detected by an automatic line search
software, FIELD \citep{Tonegawa:2015a}, which is dedicated to FMOS.

\begin{table}\begin{center}\caption{The number of galaxies
used in the RSD analysis}
\begin{tabular}{c|cccc}\hline
   &	  \multicolumn{4}{c}{line $ (S/N) > $}    \\ 
 Field    & 3.0 & 4.0 & 4.5 & 5.0  \\  
\hline
W1  & 700 & 265 & $197$ &  157 \\ 
W2  & 3125 & 1448 & $1165$ & 980 \\ 
W3  & 3191 & 1413 & $1145$ & 961 \\ 
W4  & 824 & 345 & $276$ & 222 \\
   \hline
$N_{\rm total} $ & 7840 & 3471  & 2783 & 2320 \\ 
   \hline
$f_{\rm fake}$ & 0.515 & 0.089 & 0.041 &  0.021 \\
$f_{\rm OIII} $ & 0.092 & 0.038 & 0.032 &  0.033 \\
   \hline
$N_g$ & 3530.2  & 3040.2 &  2582.4 & 2194.3
\end{tabular}\label{tab:gal} 
\begin{tabnote}
The total number $N_{\rm total}$ is the sum of the W1--4 fields. The
fake line fraction in the FastSound line emission catalog $f_{\rm
  fake}$ and the [OIII] contamination fraction in the real emission
lines $f_{\rm OIII}$ are shown.  The estimates of the number of
H$\alpha$ emission line galaxies, $N_g = N_{\rm total} (1 - f_{\rm
  fake}) (1 - f_{\rm OIII})$ is shown at the bottom of the table.  In
the FastSound emission line catalog, a fraction of galaxies appear
more than twice when they are observed more than once in overlapping
FOVs, and such duplications have been removed in this table. 
\end{tabnote}
\end{center}\end{table}

We use the FastSound catalog of emission line candidates described in
Paper II, basically assuming that the strongest line in a galaxy is
H$\alpha$.  In the case of the line threshold $S/N > 4.5$, there are
3280 galaxies in the catalog. We removed the known 52 [OIII] emitters
identified by multiple line detections as described in Paper
II (According to the statistics described in Paper II, about 10 of
these are expected to be not real [OIII] emitters, but we ignore this
small fraction).  FMOS fibers can move slightly outside the 15 arcmin
radius, but modeling the angular selection function in such regions is
not easy, and hence we removed galaxies outside the 15 arcmin radius
from each FOV, leaving 3175 galaxies.  In some of FOVs, data were not
taken in one of the two spectrographs IRS1/2 due to instrumental
troubles. In the FMOS FOV, fibers of IRS1/2 are arranged like stripes
in one direction, and IRS1 and 2 switch every two rows
\footnote{http://www.naoj.org/Observing/Instruments/FMOS/echidna.html}
\citep{Murray:2003}.
Hence if one of these is unavailable, we expect an
artificial stripe pattern of galaxy distribution in the FOV. To avoid
artificial anisotropies in the clustering analysis by this effect, we
removed such FOVs (8 in W3 and 6 in W4). This removes 153 galaxies.
We also excluded one FOV in W3 (W3\_221) from the analysis, because
the observing condition was extremely poor and emission lines were
detected only for five galaxies, which may mostly be fake emission
lines created by noise.  This cut results in a total of 3017 galaxies.

We also excluded galaxies in the regions masked by CFHTLS using the
mask data provided by the CFHTLS
team\footnote{http://www1.cadc-ccda.hia-iha.nrc-cnrc.gc.ca/community/CFHTLS-SG/docs/masks.html}.
Although the input galaxy catalog for the FastSound survey is z-band
selected, regions were removed if they are masked in the data of at
least one of the five bands, $u$, $g$, $r$, $i$, and $z$, because all
these bands are used in the SED fittings to get estimates of
photometric redshifts and H$\alpha$ fluxes in the target selection
processes. This removes about 7.8\% of galaxies, and the final galaxy
sample used in this paper becomes 2783 galaxies.  In appendix
\ref{sec:sys_cfhtls} we test how the final cosmological results are
affected by the treatment of masked regions.

\begin{figure}\begin{center}\FigureFile(70mm,70mm){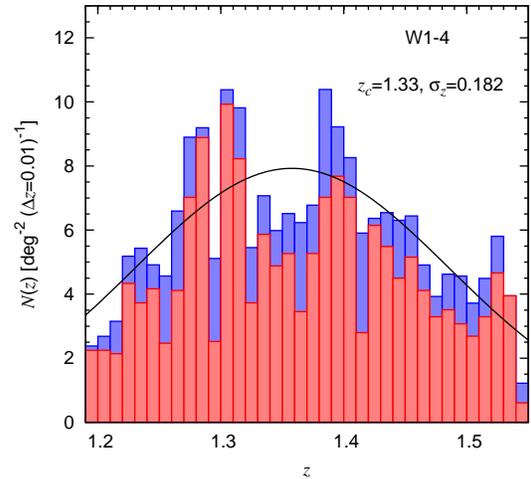}
\end{center}
 \caption{Redshift distribution of our galaxy sample (the
   signal-to-noise ratio threshold $S/N>4.5$ for emission lines),
   combined for all the four fields W1-4 ({\it red histogram}). The blue
   histogram is the distribution after correcting for the decrease of
   the detection efficiency by the OH masks of FMOS.  The black solid line is
   our best-fitting model to the blue histogram , which is used as the
   reference for the distribution without clustering to calculate
   correlation functions.}
\label{fig:zdist}\end{figure}

We constructed galaxy samples with several different thresholds for
the line signal-to-noise.  The number of galaxies in our samples are
summarized in table \ref{tab:gal}.  The total number of galaxies in
the four fields is given as $N_{\rm total}$, but not all of the
emission lines in the catalog are real.  As discussed in Paper I, 
the fraction of fake lines due to noise can be estimated by statistics of inverted frames. 
The inverted frames are obtained by exchanging object and sky frames in the reduction process, 
and the fake lines are found to be independent of the observed wavelength (see figure 9 of Paper I). 
The estimated fake line fraction $f_{\rm fake}$ is also
given in the table, and $f_{\rm fake}$ is higher for lower line $S/N$.
Even if an emission line is real, it may not be H$\alpha$.  The
contamination fraction of other lines was studied in detail in Paper
II, based on galaxies with multiple lines and the stacked spectrum
of FastSound galaxies.
We found that the major contaminant is the [OIII] doublet. The
estimated contamination fraction $f_{\rm OIII}$ in all real lines is
also shown in the table. Then the number of galaxies with real
H$\alpha$ emission lines can be estimated as $N_g = N_{\rm total} (1 -
f_{\rm fake}) (1 - f_{\rm OIII})$. In this work, we use the sample
with $S/N>4.5$ in our baseline analysis, with $f_{\rm fake} = 0.041$
and $f_{\rm OIII} = 0.032$.  The effect of the fake and non-H$\alpha$
lines will be corrected in the clustering analysis (see section
\ref{sec:blunder} below).  We will use the samples with the other
thresholds, $S/N>4.0$ and $S/N>5.0$, for a test of systematic errors.

\begin{figure}
\begin{center}\FigureFile(82mm,82mm){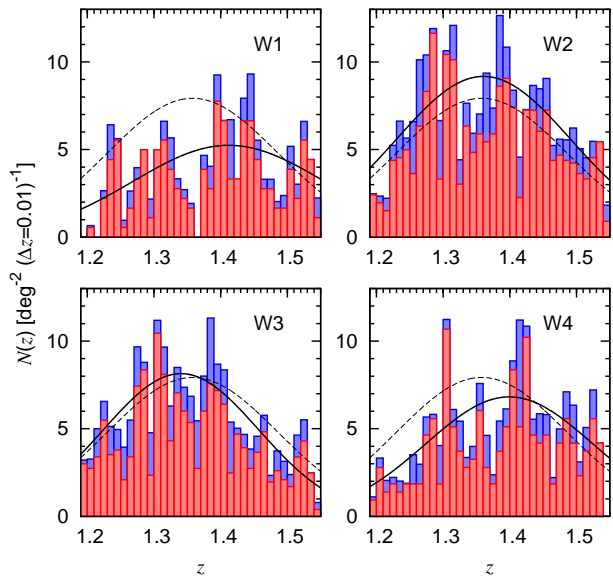} \end{center}
 \caption{Same as figure \ref{fig:zdist},
but for each of the sub-fields of CFHTLS W1--4.
The dashed lines are the best fitting model to the whole sample W1-4 as a reference (the same as the solid line in figure \ref{fig:zdist}),
while the solid line is the best fitting model to each sample with the parameters $(z_c,\sigma_z)=(1.38,0.206)$, $(1.33,0.187)$, $(1.32,0.166)$ and $(1.38,0.179)$ for 
W1-W4, respectively.}
\label{fig:zdist_w1-4}\end{figure}

\begin{figure} 
\begin{center}\FigureFile(83mm,83mm){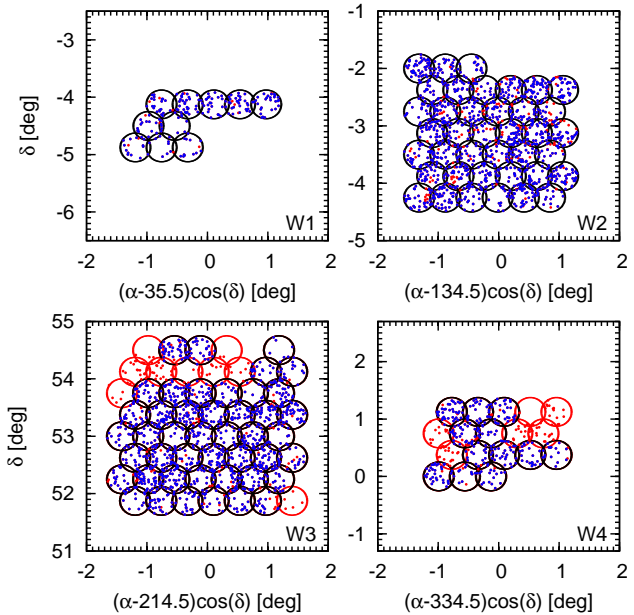}\end{center} 
\caption{Angular distribution of the galaxies with line $S/N>4.5$ for the
  four fields of CFHTLS W1--4. The blue points are galaxies used for
  our cosmological analysis while the red ones are those excluded by
  our selection criteria described in section
  \ref{sec:fastsound}.} \label{fig:ntarget_cfhtlsw1-4_v1p70}
\end{figure}

The redshift distributions of the galaxy sample ($S/N > 4.5$) are
shown in Figs. \ref{fig:zdist} (total) and \ref{fig:zdist_w1-4} (for
each of W1--4).  The mean redshift of the sample is $\bar{z}=1.36$.
The angular positions of galaxies are shown for the four fields in
Fig.  \ref{fig:ntarget_cfhtlsw1-4_v1p70}.  
The effect that the observing conditions are different for each FOV, 
which would make the galaxy sample inhomogeneous, will be 
discussed in appendix \ref{sec:sys_selection}.

\subsection{Random catalog}\label{sec:random}
Constructing a random catalog, i.e., a virtual galaxy catalog without
clustering, is one of the most important tasks to measure reliable
correlation functions from observed data. In the
following we describe how we construct the radial and angular
selection functions to create our random catalog.

\subsubsection{Angular selection function}\label{sec:angular}

Our observation was not made uniformly for angular fields, and the
detection efficiency for each field-of-view, i.e., the angular
selection function, needs to be estimated.  For the angular selection
we consider a simple model that uses the ratio of the number of
galaxies detected in our sample to that of photometrically selected
targets for spectroscopy: $W_A=N_A^{\rm det}/N_A^{\rm tar}$ for a FOV A.  This
takes into account the fluctuation of the number density of target
galaxies in contrast to the fixed number of fibers in a FOV, and also
the line detection efficiency for galaxies to which FMOS fibers were
allocated, which should depend on observing conditions.

Circular FOVs of FastSound are arranged by a hexagonal tiling, and
there are overlapping regions. For non-overlapping regions, we simply
estimate the selection function by the above formula. However, in the
overlapping region between FOVs A and B, some galaxies are detected twice. 
The galaxies which are detected at least once in the two observations
are included in the catalog, thus a special treatment is required for the overlapping region. 
It is difficult to simply
adopt the above formula ($N^{\rm det}/N^{\rm tar}$) in these regions because
of the small number of detected emission line galaxies in overlapped
regions and hence large Poisson uncertainties. Therefore we first estimate
the detection efficiency for fiber-allocated galaxies in FOV A as $w_A
\equiv N_A^{\rm det}/N_A^{\rm fib}$, where $N_A^{\rm fib}$ is the number of
galaxies to which FMOS fibers are allocated in FOV A. 
The typical values of $w_A$ are 0.08 and 0.10 for the regions where 
fibers of IRS1 and IRS2 are assigned, respectively, and the value for 
the total area is 0.09. When galaxies
are observed in both FOVs A and B, we assume that the detection
efficiency is determined by the FOV having a better
efficiency.  Therefore, if we define the FOV A so that it has a better
efficiency ($w_A > w_B$), the selection function is calculated as
\begin{equation}
W_{A \cap B} = \frac{ w_A (N_A + N_{AB}) + w_B N_B }{N^{\rm tar}_{A \cap B}} \label{eq:angular_overlap}
\end{equation}
where $N_A$ ($N_B$) is the number of galaxies in the overlapping
region which were observed only in the observation of FOV A (B),
$N_{AB}$ the number of galaxies which were observed in both the
observations, and $N^{\rm tar}_{A \cap B}$ the number of target galaxies
in the overlapping region.

\subsubsection{Radial selection function}\label{sec:radial}
To construct the radial selection function, we adopt a commonly-used
functional form \citep{Baugh:1993}, as 
\be n(z)=Az^2\exp{\left[
  -\frac{(z-z_c)^2}{\sigma_z^2} \right]}, \label{eq:radial_3para} 
 \ee
where $A$, $z_c$, and $\sigma_z$ are free parameters.  We have also
tried another commonly-used functional form (e.g.,
\cite{de-la-Torre:2013}), \be n(z)=\left(\frac{z}{z_0}\right)^\alpha
\exp{\left(\frac{z}{z_0}\right)^\beta},\ee where $\alpha$, $\beta$,
and $z_0$ are free parameters.  We have confirmed that the difference
using the two forms is negligibly small for the final cosmological
results, and we will present the best-fit parameter set only for
equation (\ref{eq:radial_3para}) below.

\begin{figure}\begin{center}\FigureFile(70mm,70mm){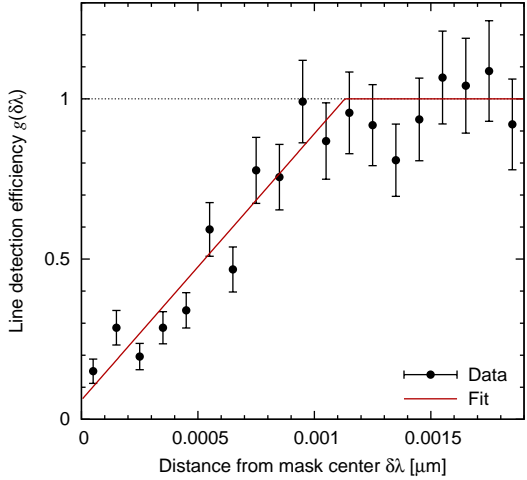} \end{center}
 \caption{The ratio of the number of observed galaxies and mock galaxies as a function of the distance from the center of OH masks. The solid line 
 is a fitting function given by equation (\ref{eq:efficiency_loss}).}
\label{fig:ohmask}
\end{figure}

As described in Paper I, FMOS uses OH masks to suppress the strong OH
airglow emission lines in the NIR bands, and this effect should
be treated carefully.  We examined the decrease of detection
efficiency around the masks by plotting the number of detected
galaxies as a function of distance of emission line wavelength from
the nearest mask center, $\delta \lambda = | \lambda_{\rm
  line}-\lambda_{\rm mask} |$, in comparison with that of a mock galaxy
sample expected when there are no masks.
The decrease of detection efficiency is fit by a simple
function as 
\bey g(\delta\lambda)= \left\{ \begin{array}{ll} 831.5 \;
  \delta\lambda + 0.06, & (\delta\lambda \leq 0.00113 \ \mu {\rm m})
  \\ 1 & (\delta\lambda > 0.00113 \ \mu {\rm m}) \ . \\
\end{array} \right. \label{eq:efficiency_loss}
\eey
The width of a mask is 0.0007 $\mu$m, i.e., 
0.00035 $\mu$m from the mask center to the mask borders. 
The efficiency decreases to less than 10\% at the 
mask centers. Figure \ref{fig:ohmask} shows the comparison of 
the observed decrease of detection efficiency to the fitting function. 

We applied this formula to correct the redshift histogram shown in
Figs. \ref{fig:zdist} and \ref{fig:zdist_w1-4}, and the corrected
histogram is indicated by blue. During the period of the FastSound
observing runs, the OH masks of IRS1 was updated in July 2012, and
after that the mask patterns of IRS1 and 2 became exactly the
same. This is taken into account in the histogram correction.  Then
the model function of the radial selection function [equation
  (\ref{eq:radial_3para})] was fit to the corrected redshift
histograms, and we found the best-fit values of
$(z_c,\sigma_z)=(1.34,0.179)$. This fitted function is shown in
Fig. \ref{fig:zdist}. We also performed this fit to each field of
W1--W4, and show the best fit models in Fig. \ref{fig:zdist_w1-4} 
as the solid lines.

We then construct the random catalog by applying equation
(\ref{eq:radial_3para}), taking into account the effect of the
detection efficiency decrease by OH masks again using equation
(\ref{eq:efficiency_loss}).  In our baseline analysis, we use the
radial selection functions fitted to each of W1--W4 fields separately.
In appendix \ref{sec:systematics} we will examine how the final
cosmological results change when we use the radial selection function
determined using the combination of the four fields.

\section{Correlation function measurement}\label{sec:measurement}

In this section we measure the redshift-space correlation function of the FastSound galaxy sample $\xi^s({\bf r})$. 
For this purpose we count the galaxy pairs in bins of comoving separation.
We can choose the separation vector between two galaxies ${\bf r}$ arbitrarily, such as ${\bf r}=(r)$, $(r_p,r_\pi)$, $(r,\mu_{\vecr})$, and so on,
where $r_p$ and $r_\pi$ are respectively the separations perpendicular and parallel to the line of sight, 
$r=\sqrt{r_p^2+r_\pi^2}$ and $\mu_\vecr=r_\pi/r$ is the direction cosine in configuration space. 
We adopt the Landy-Szalay estimator \citep{Landy:1993} to measure the correlation function, 
\be
\xi^s({\bf r})=\frac{DD-2DR+RR}{RR}, \label{eq:ls}
\ee
where $DD$, $RR$ and $DR$ are the normalized counts of galaxy-galaxy, random-random and galaxy-random pairs, respectively, and 
the superscript $s$ denotes the quantity defined in redshift space.
To compute $DR$ and $RR$ we generate a random catalog $55$ times denser than our galaxy catalog using the radial and angular selection functions described in section \ref{sec:random}.

Two flat cosmological models, $\Omega_{m}=1-\Omega_\Lambda=0.317$ \citep{Planck-Collaboration:2014} and $\Omega_{m}=1-\Omega_\Lambda=0.270$ \citep{Hinshaw:2013} are considered when redshift is converted to the radial distance for each galaxy. 
Since the two cases give almost the same cosmological results, we will mainly show the results with the case of WMAP ($\Omega_{m}=0.270$) otherwise stated.

\subsection{Accounting for fiber allocation failures}\label{sec:fiber_alloc}

\begin{figure}\begin{center}\FigureFile(75mm,75mm){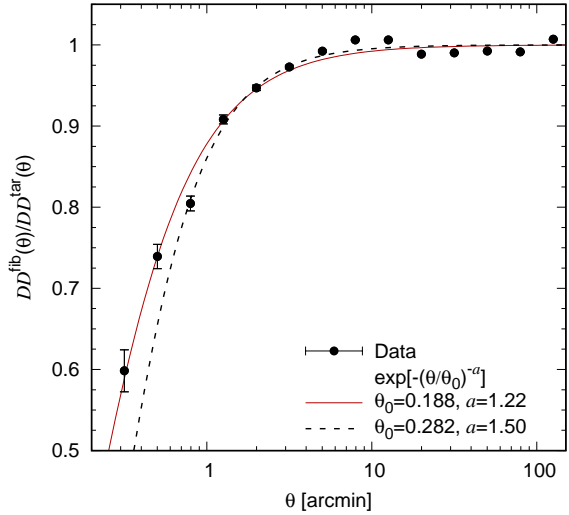} \end{center}
 \caption{ The ratio of the two normalized galaxy-galaxy pair counts,
   $DD^{\rm fib}(\theta)/DD^{\rm tar}(\theta)$ where
   $DD^{\rm fib}$ is for the sample to which fibers were assigned and
   $DD^{\rm tar}$ is for galaxies in the FastSound target catalog.
   The red sold line is the best fitting model fitted for
   $0.3<\theta<50$ [arcmin] with the form
   $\exp{[-(\theta/\theta_0)^{-a}]}$, while the black dashed line is
   for $0.7<\theta<50$ [arcmin].}
\label{fig:fiberalloc}
\end{figure}

In highly clustered regions, some of the targets are not assigned
fibers due to their finite number.  We call this effect fiber
allocation failures, and it affects the correlation function in a
similar manner to the fiber collision effect due to the finite
physical sizes of fibers \citep{Lin:1996,Hawkins:2003}.  We correct
for this effect by using the ratio of galaxy pair counts as a function
of separation angle, $DD^{\rm fib}(\theta)/DD^{\rm tar}(\theta)$,
where $DD^{\rm fib}(\theta)$ is the normalized pair count of the
FastSound galaxies to which fibers were assigned, and $DD^{\rm
  tar}(\theta)$ is for galaxies in the FastSound target catalog before
fiber allocation. Figure \ref{fig:fiberalloc} shows $DD^{\rm
  fib}(\theta)/DD^{\rm tar}(\theta)$.  The deviation of the ratio from
unity is due to the failure of allocating fibers to galaxies in
clustered regions, thus this effect can be corrected for by using a
weight equal to the inverse of this ratio to the data-data pair count
(see e.g. \cite{Hawkins:2003}). One can thus see that the fiber
allocation correction starts to be important at an angular scale of
$\theta \sim 1 $ arcmin. We found that the ratio can be well described
by a simple function, \be DD^{\rm fib}(\theta)/DD^{\rm tar}(\theta) =
\exp{\left[-(\theta/\theta_0)^{-a}\right]}
\label{eq:fiberalloc},\label{eq:fiber_alloc}\ee
where $\theta_0$ and $a$ are free parameters.  If we adopt the
simplest estimator for the angular correlation function,
$w(\theta)=DD/RR-1$, the ratio in equation (\ref{eq:fiber_alloc}) is
equal to $[1+w^{\rm fib}(\theta)]/[1+w^{\rm tar}(\theta)]$
\citep{Hawkins:2003}.  The red solid line in figure
\ref{fig:fiberalloc} shows the best fitting model with
$(\theta_0,a)=(0.188,1.22)$ determined at the angular scales
$0.3<\theta<50$ [arcmin].  We will weight the pair count for
data-data, $DD$ in equation (\ref{eq:ls}), by the inverse of equation
(\ref{eq:fiber_alloc}).  The result when the fitting range is changed
as well as the result when the effect of this fiber allocation
correction is ignored will be tested in appendix
\ref{sec:sys_fiber_alloc}.

\subsection{Integral constraint}\label{sec:ic}
To take into account the finite survey volume the integral constraint needs to be added to the measured correlation function [equation (\ref{eq:ls})].
It is computed as \citep{Peebles:1976},
\be
I.C. = \frac{\sum_{s_i}\xi^s(s_i)RR(s_i)}{\sum_{s_i}RR(s_i)}, \label{eq:ic}
\ee
where we can compute $\xi^s$ theoretically and we simply use linear theory for it (see section \ref{sec:basics} below). 
Since we use our correlation function measurement maximally up to $r_{\rm max}=80\himpc$, this effect is negligibly small, 
nevertheless we add this constraint to our measurement. 

\begin{figure}\begin{center}\FigureFile(83mm,83mm){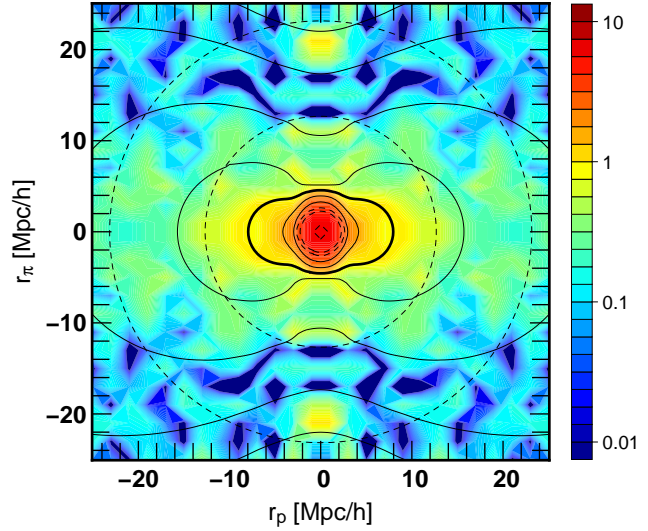} \end{center}
 \caption{Anisotropic correlation function $\xi^s(r_p,r_\pi)$ measured from the FastSound galaxy catalog shown as the color contours  in cells of side $\Delta r_p=\Delta r_\pi=2\himpc$. 
 The solid contours are the best-fitting model of the linear Kaiser plus nonlinear matter power spectrum with $(f\sigma_8,b\sigma_8)=(0.478,0.818)$ and the redshift blunder fraction $f_{\rm blund}=0.071$ described in section \ref{sec:analysis}. The thick line corresponds to $\xi^s=1$, and the value of $\xi^s$ changes logarithmically with 0.5 ($\Delta \log_{10}\xi^s=0.5 $), decreasing outward. 
Note that the feature seen on small scales along the line of sight in the theoretical model, $r_p\sim 0\himpc$ and $r_\pi<5\himpc$, is due to the numerical effect of the Fourier transform but the data at such scales are not used in the analysis.
The dashed, circular lines are the monopole correlation function $\xi^s_0(r)$ computed using the same parameter set of $(f\sigma_8,b\sigma_8,f_{\rm blund})$ to clarify the effect of the anisotropy by RSD. Same as the RSD model, the inner and outer contours correspond to $\log_{10}\xi^s=-0.5$ and $-1$, respectively.
 }
\label{fig:2dxi}
\end{figure}

\subsection{Correlation function of the FastSound galaxies}
First we measure the correlation function as a function of comoving separation perpendicular and parallel to the line of sight, $\xi^s(r_p,r_\pi)$ \citep{Peebles:1980,Davis:1983}, at $z\sim 1.4$.
The result is plotted in figure \ref{fig:2dxi} as color contours. 
 We choose the size of square bins to be $\Delta r_p=\Delta r_\pi=2\himpc$. One can see that the iso-correlation contours are squashed along the line of sight, as expected from the RSD signal, particularly at the scale around $10<r<20\himpc$ (for the anisotropic contours obtained at lower redshifts, see e.g., \cite{Peacock:2001,Ross:2007,Guzzo:2008,de-la-Torre:2013,Guo:2015a}). 
On the other hand, nonlinear velocity dispersion including the Finger-of-God effect \citep{Jackson:1972} that elongates the clustering signal along the line of sight is not prominent except on very small scales $r<5\himpc$. We will not use these scales for the following cosmological analysis. This is expected given that our spectroscopic targets are emission line galaxies preferentially avoid high density regions, reside in small dark matter halos and are predominantly central galaxies (see e.g., \cite{Koda:2015}). 

The real-space correlation function can be estimated from the anisotropic correlation function $\xi^s(r_p,r_\pi)$ by projecting it along the line of sight and eliminating RSD, $w_p(r_p)=\int dr_\pi \xi^s(r_p,r_\pi)$. Although the projected correlation function plays an important role for modeling the halo occupation distribution, we do not discuss the real-space clustering and it will be presented in \hikaget  in detail.

\begin{figure}\begin{center}
\FigureFile(79mm,79mm){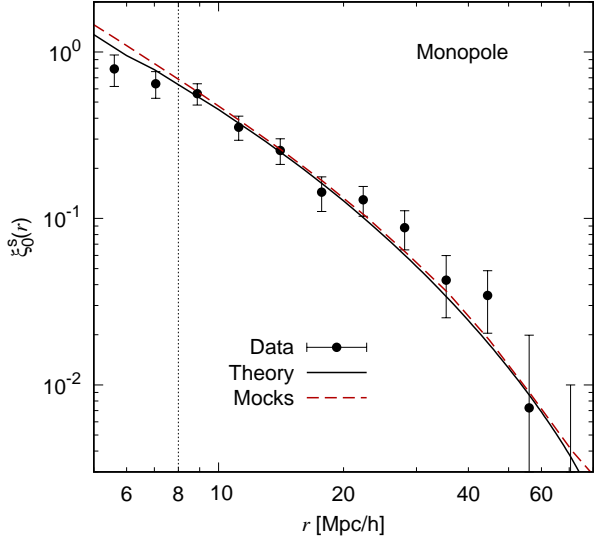}
 \end{center}
 \caption{Monopole correlation function of the FastSound galaxy sample.
 The points are the measurements from the FastSound galaxy sample, and the error bars are from the mock catalogs. 
 The black solid lines are the best-fitting model of the linear Kaiser plus the nonlinear matter power spectrum with $(f\sigma_8,b\sigma_8,f_{\rm blund})=(0.478 ,0.818,0.084)$ (see section \ref{sec:analysis}).
 The vertical lines are the minimum scale we used for the cosmological fits, $r_{\rm min}=8\himpc$.
  The red dashed lines are the results from our mock catalogs averaged over the whole 640 realizations.
}
\label{fig:mono}
\end{figure}

\begin{figure}\begin{center}
\FigureFile(79mm,79mm){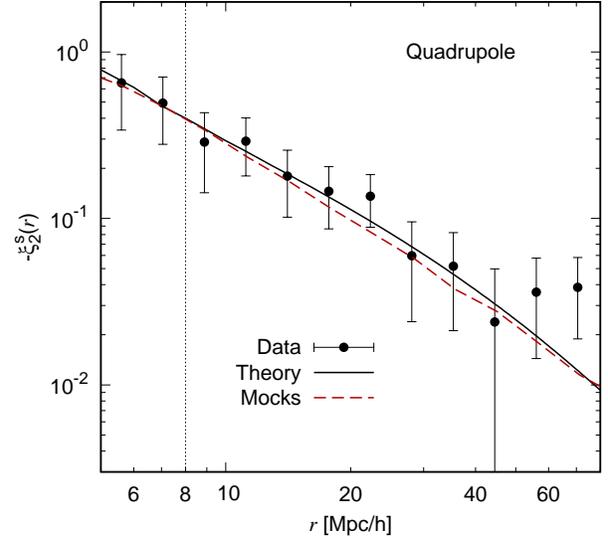}
 \end{center}
 \caption{Same as figure \ref{fig:mono} but for the quadrupole correlation function of the FastSound sample.
  Note that the quadrupole is a negative quantity for all the scales probed here, thus we plot $-\xi^s_2$.
}
\label{fig:quad}
\end{figure}

Next we present the two lowest multipoles of the correlation function, e.g., monopole and quadrupole. 
They can be measured by integrating the anisotropic correlation function multiplied by the Legendre polynomials ${\cal L}_\ell$ over angle, as 
\citep{Hamilton:1992}
\bey
\xi_\ell^s(r)&=&\frac{2\ell+1}{2}\int^{1}_{-1} \xi^s({\bf r}){\cal L}_\ell(\mu_\vecr) d\mu_\vecr, \label{eq:xi_multi2}\\
\xi^s(\vecr) &=& \sum_{l=0} \xi_\ell^s(r){\cal L}_\ell(\mu_\vecr). \label{eq:xi_multi}
\eey
To do this, we estimate the full 2D correlation function $\xi^s(r,\mu_\vecr)$ and perform the sum over $\mu_\vecr$. We adopt the bin size to be $\Delta \log_{10}{r}=0.1$ and $\Delta\mu_\vecr=0.1$. 
The resulting monopole and quadrupole correlation functions are shown as the black points in figures \ref{fig:mono} and \ref{fig:quad}, respectively. 
We detect non-zero quadrupole signals on scales $r>5\himpc$.
Note that the clustering amplitude is reduced compared to the true clustering due to the presence of redshift blunders discussed in section \ref{sec:fastsound}. 
We will take into account this effect in theoretical modeling of the correlation function in section \ref{sec:blunder} below.
To reduce the degree of freedom, we analyze the multipoles for cosmological constraints rather than the full 2D correlation function.

\begin{figure}\begin{center}
\FigureFile(80mm,80mm){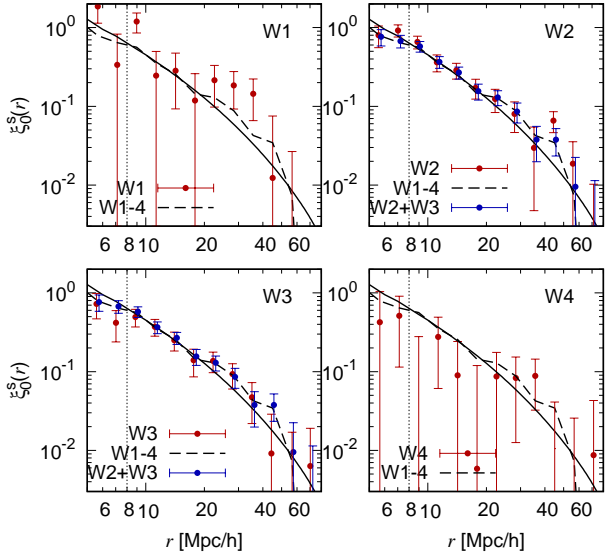}
 \end{center}
 \caption{Monopole correlation function of each of the 4 CFHTLS fields, W1-4. 
 The error bars are from the mock catalogs of each of the fields. The monopole correlation of the whole 4 fields is shown as the dashed lines
  (the same as the points in figure \ref{fig:mono}).
 The black solid lines are the best-fitting model of the linear Kaiser plus the nonlinear matter power spectrum for the whole samples (the same as the solid line in figure \ref{fig:mono}).
 For the panels of W2 and W3 results, we also plot the measurement from the combined sample of W2 and W3 fields as the blue points.
}
\label{fig:mono_w1-4}
\end{figure}

\begin{figure}\begin{center}
\FigureFile(80mm,80mm){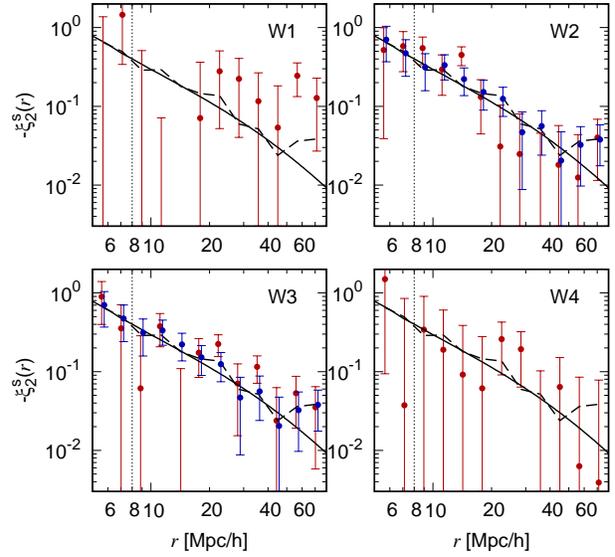}
 \end{center}
 \caption{Same as figure \ref{fig:mono_w1-4} but for the quadrupole correlation function for each of the 4 fields.
}
\label{fig:quad_w1-4}
\end{figure}

We present the monopole and quadrupole correlation functions measured from each of the four CFHTLS survey fields in figures \ref{fig:mono_w1-4} and \ref{fig:quad_w1-4}, respectively.
For comparison the measurement from the total 4 fields (the points in figures \ref{fig:mono} and \ref{fig:quad}) is plotted as the dashed lines. 
The correlation functions measured from the fields W1 and W4 are very noisy because these samples are about 4 times smaller than the W2 and W3 fields, but the measurement from each field is largely consistent with each other. At the panels of the results for W2 and W3, we also show the correlation functions from the combined sample of W2 and W3 as the blue points. 
The measurements are very close to the dashed lines, which implies that the most of the contribution is coming from the largest two fields, W2 and W3 fields. 

\subsection{Covariance matrix}\label{sec:covariance}

In order to estimate the covariance matrix for the measured correlation function, we use outputs of $N$-body simulations run by \citet{Ishikawa:2014}. 
They adopted the best-fitting cosmological parameters in the WMAP 7-year data \citep{Komatsu:2011} and created 40 realizations each of which contains $1024^3$ dark matter particles in a cubic box of side $700\himpc$. 
Since the volume of the FastSound survey is much smaller than each simulation box, from the 40 realizations we create 640 mock halo catalogs that have survey geometry the same as our W1-4 fields. Galaxies are populated into the halos according to the halo occupation distribution (HOD) model \citep{Cooray:2002, Zheng:2005} constrained by the real-space clustering of the FastSound galaxies (\hikage). We find that the observed clustering can be well modeled without populating satellite galaxies, i.e., using only by 2-halo term, consistent with the recent finding of \citet{Koda:2015}.
The best-fitting HOD model is the Gaussian distribution of central galaxies with the average mass of $M_c=4.7\times 10^{12}M_\odot /h$ with the scatter $\sigma_{\log{M}}=0.4$.

We compute the correlation functions from each of the mock catalogs, $\xi_{\ell,k}^s$ $(\ell=0,2)$, in the same way as our observation, and estimate the covariance matrix as 
\bey
  \hat{C}_{ij} &=& \frac{1}{N_{\rm mock}-1} \nn \\
&& \times \sum^{N_{\rm mock}}_{k=1}\left[\xi^s_{\ell,k}(r_i)-\bar{\xi}^s_\ell(r_i)\right]
\left[\xi^s_{\ell,k}(r_j)-\bar{\xi}^s_\ell(r_j)\right],  \label{eq:cov}
\eey
where $N_{\rm mock}=640$ and $\bar{\xi}^s_\ell = \frac{1}{N_{\rm mock}}\sum_{k=1}^{N_{\rm mock}} \xi^s_{\ell,k}$.
Let the number of bins for the monopole and quadrupole $N_{\rm bin}$, then the covariance becomes $2N_{\rm bin}\times 2N_{\rm bin}$ matrix. 

\begin{figure}\begin{center}\FigureFile(85mm,85mm){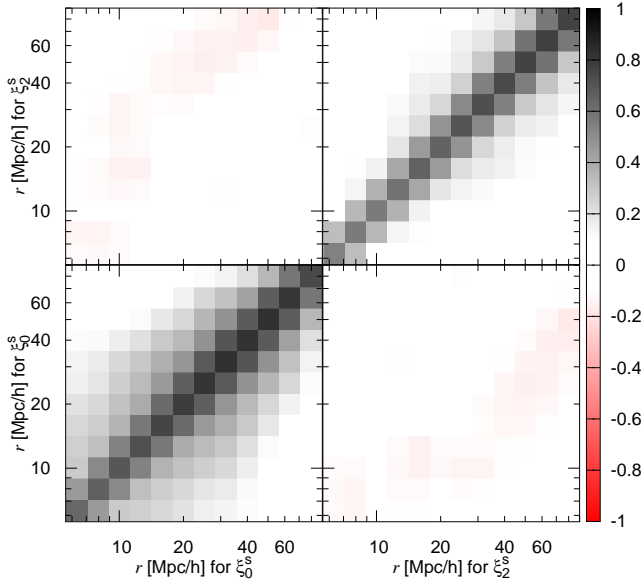} \end{center}
 \caption{Correlation matrix for the multipole and quadrupole correlation functions estimated from 640 mock galaxy subsamples. 
}
\label{fig:cov}
\end{figure}

The underestimation of the covariance due to the finite number of realizations is corrected for following \citet{Hartlap:2007},
\be
C^{-1}=\frac{N_{\rm mock}-N_{\rm bin}-2}{N_{\rm mock}-1} \hat{C}^{-1}. \label{eq:cov2}
\ee
The covariance matrix normalized by the diagonal components, namely the correlation matrix $C_{ij}/(C_{ii}C_{jj})^{1/2}$, is shown in figure \ref{fig:cov}.
As a reference we show the matrix for the scales $5<r<90\himpc$, wider than the scales we will use in the following cosmological analysis. The order of the matrix used in the following analysis will be smaller than the presented one.

The error bars for the multipoles in figures \ref{fig:mono} and \ref{fig:quad} are the diagonal parts of the covariance matrix $\sigma_i=C_{ii}^{1/2}$. 
The multipole correlation functions averaged over all the mock results $\bar{\xi}^s_\ell$ are plotted as the red dashed lines in figures \ref{fig:mono} and \ref{fig:quad}.

\section{Theory}\label{sec:basics}

\subsection{Models of redshift-space correlation function}

In this section we review the theoretical models of the redshift-space correlation function which will be considered for our cosmological analysis. 
Under the plane-parallel approximation, the redshift-space power spectrum with the most general form can be written as \citep{Scoccimarro:1999,Scoccimarro:2004}
\bey
P^s(\veck)&=&\int \frac{d^3 {\bf r}}{(2\pi)^3} e^{i {\bf k}\cdot{\bf r}}\left\langle e^{-ifk\mu_\veck [u_z(\vecx)-u_z(\vecx')]}\right. \nn \\
&&\times \left. [\delta ({\bf x}) +f\nabla u_z({\bf x}) ][\delta ({\bf x}') +f\nabla_z u_z({\bf x}') ] \right \rangle, \label{eq:scoccimarro1999}
\eey
where $\mu_{\bf k}$ is the direction cosine between the observer's line of sight and the wavevector ${\bf k}$, $k=|{\bf k}|$, 
$\vecr = \vecx-\vecx'$, $u_z(\vecx)=-v_z(\vecx)/(aHf)$, $\Delta u_z=u_z(\vecx)-u_z(\vecx')$, and $v_z$ is the radial component of the velocity field.
Just like the case for the correlation function [equation (\ref{eq:xi_multi})], one can consider the multipole expansion for the power spectrum, 
\be
P^s_\ell(k) = \frac{2l+1}{2} \int^{1}_{-1} P^s(\veck){\cal L}_\ell(\mu_{\bf k}) d\mu_\veck.
\ee
The multipole moments of the power spectrum are related to those of the correlation function through the relation,
\be
\xi_\ell^s(r)=i^l\int \frac{dkk^2}{2\pi^2}P_\ell^s(k)j_\ell(kr),
\ee
where $j_\ell$ is the $\ell$-th order spherical Bessel function. 
In this way our theoretical model of the correlation function can be compared to the measurement [equation (\ref{eq:xi_multi2})]. 

The logarithmic derivative of the linear growth rate, $f$, is sensitive to modification of gravity models.
That is the reason why a detailed analysis of the growth of structure through RSD enables us to directly test gravity on large scales.

Although equation (\ref{eq:scoccimarro1999}) is a general form and contains the full nonlinear properties, such as nonlinear gravity and velocity dispersion, it is hard to compute the full equation and we need to consider approximation to extract cosmological information from the observed galaxy clustering. 

\subsubsection{Linear RSD}
The simplest and most well-known approximation is linear theory of density perturbation. Under the assumption, the above equation is greatly simplified \citep{Kaiser:1987}, 
\be
P^s({\bf k})=(b+f\mu_{\bf k}^2)^2 P_{\delta\delta}^{\rm lin}(k), \label{eq:kaiser}
\ee
where $b$ is the linear bias parameter and $P_{\delta\delta}^{\rm lin}(k)$ is the density power spectrum of dark matter in linear theory. 
In linear theory the expansion has only three components, $P_0$, $P_2$ and $P_4$, and they can be described as
\bey
P_0^s(k)&=&\left( b^2+\frac{2}{3}bf + \frac{1}{5}f^2\right)P_{\delta\delta}^{\rm lin}(k), \label{eq:p0}\\
P_2^s(k)&=&\left( \frac{4}{3}bf+\frac{4}{7}f^2 \right)P_{\delta\delta}^{\rm lin}(k), \label{eq:p2}\\
P_4^s(k)&=&\frac{8}{35}f^2P_{\delta\delta}^{\rm lin}(k). \label{eq:p4}
\eey
We use the CAMB code \citep{Lewis:2000} to compute the linear power spectrum $P_{\delta\delta}^{\rm lin}$.
Because $P_{\delta\delta}^{\rm lin}\propto \sigma_8^2$, free parameters for the linear RSD are $f\sigma_8$ and $b\sigma_8$ when fundamental cosmological parameters are fixed, and there is a strong degeneracy between the two parameters.
As is clear from the form of RSD in the 2D power spectrum (equation \ref{eq:kaiser}) and its multipoles (equations \ref{eq:p0}-\ref{eq:p4}), 
the determination of the correlation amplitude constrains the combination, $bf\sigma_8^2$.
On the other hand, large-scale anisotropies caused by RSD constrains the ratio of the two parameters, 
$\beta=fb^{-1}$.
The monopole better determines the bias, while 
the growth rate $f$ can be better determined by the high-order multipoles. 
Particularly, the hexadecapole is a bias-free estimator for $f\sigma_8$, but its measurement is noisier than the monopole or quadrupole. 
Thus, analyzing the full redshift-space clustering of galaxies allows one to break the degeneracy and 
determine the parameter $f$ we are interested in. 

\subsubsection{Non-linear RSD}

It is well known that on small scales various nonlinear effects alter the shape of the power spectrum (nonlinear gravity, velocity dispersion and biasing), so analyzing data with the linear Kaiser formula could potentially bias the cosmological constraints (e.g., \cite{Scoccimarro:1999,Scoccimarro:2004,Tinker:2006,Okumura:2011,Jennings:2011,Kwan:2012,Bianchi:2012}). 

We consider two commonly-used models \citep{Peacock:1994,Park:1994,Fisher:1995,Scoccimarro:2004,Okumura:2012}, 
\bey
P^s({\bf k})& =& \left[ b^2 P_{\delta\delta}(k) +2bf \mu^2 P_{\delta\theta}(k) + f^2 \mu^4 P_{\theta\theta}(k) \right] \nn \\
&& \ \ \ \ \ \ \ \ \ \ \ \ \ \ \ \ \ \ \ \ \ \ \ \ \ \ \ \ \ \ \ \ \ \ \ \ \ \ \ \ \ \ \ \ \ \ \ \ \ \ \times G^2(k\mu\sigma_{\rm v}), \label{eq:scoccimarro}\\
P^s({\bf k})&=&(b+f\mu_{\bf k}^2)^2 P_{\delta\delta}(k) G^2(k\mu\sigma_{\rm v}), \label{eq:nl_kaiser}
\eey
where $P_{\delta\delta}$, $P_{\delta\theta}$ and $P_{\theta\theta}$ are the nonlinear, auto power spectrum of density, that of velocity divergence and their cross power spectrum, respectively. The function $G$ is the damping function by the nonlinear velocity dispersion, and the Gaussian and Lorentzian functions are commonly chosen to represent this damping. 
Because the two functions are equivalent at the lowest order, $G^2(x) \simeq 1-x^2$ and we are interested only in the large-scale clustering at high redshift, we consider only the Gaussian function,
$G(k\mu\sigma_{\rm v})=\exp{[-(k\mu \sigma_{\rm v})^2/2]}$,
where $\sigma_{\rm v}$ is the nonlinear velocity dispersion parameter. 
Equation (\ref{eq:nl_kaiser}) can be obtained if we assume the linear relation between the density and velocity fields, $P_{\delta\delta}=P_{\delta\theta}=P_{\theta\theta}$, in equation (\ref{eq:scoccimarro}).
For these models, free parameters are $f\sigma_8$ and $b\sigma_8$, the same as the linear RSD model, and $\sigma_{\rm v}$. 
For each of the two models, we also consider the case when the velocity dispersion is fixed to $\sigma_{\rm v}=0$, namely $G=1$, thus 
totally four nonlinear models will be tested.
In the case of $G=1$, the multipole expression for equation (\ref{eq:nl_kaiser}) can be obtained by replacing $P_{\delta\delta}^{\rm lin}$ by the nonlinear power $P_{\delta\delta}$ in equations (\ref{eq:p0}) - (\ref{eq:p4}), and that for equation ({\ref{eq:scoccimarro}) by replacing them by $P_{\delta\delta}$, $P_{\delta\theta}$ and $P_{\theta\theta}$ in equations (\ref{eq:p0}), (\ref{eq:p2}) and (\ref{eq:p4}), respectively. For the models with the Gaussian velocity dispersion, we can still express the multipoles analytically with the compact forms [see equations (4.9)-(4.13) of \cite{Taruya:2009}]. 
Under the assumption of linear theory of density perturbation, the above two models converge to the linear Kaiser formula [equation (\ref{eq:kaiser})].

We use the improved version of HALOFIT \citep{Smith:2003} developed by \citet{Takahashi:2012} to compute the nonlinear matter power spectrum $P_{\delta\delta}$. 
Moreover, the power spectra of velocity divergence, $P_{\delta\theta}$ and $P_{\theta\theta}$ are computed using the fitting formulae derived by \citet{Jennings:2012}. 
Since the formulae break down at the large-scale limit, $k<0.006\hmpci $ , we replace them by the linear theory prediction at such $k$. 
We will test whether the 5 models can give unbiased constraints for our FastSound galaxy sample using the $N$-body simulation data in section \ref{sec:mock_test}.
Note also that modeling the galaxy bias is another important issue to use RSD for precision cosmology because the galaxy bias is known to suffer from nonlinearity (See, e.g., \citet{Saito:2014} for one of the latest studies to formulate the nonlinear bias). 
However, given our data quality, we will consider only the linear, constant bias. We will also test the assumption using the mock galaxy sample in section \ref{sec:mock_test}.

There are numerous models for nonlinear RSD of biased objects based on a halo model, perturbation theory and $N$-body simulations 
(e.g., \cite{Seljak:2001,White:2001,TInker:2007,Matsubara:2008a,Taruya:2010,Reid:2011,Nishimichi:2011,Okumura:2012b,Vlah:2013,Song:2014a,Kitaura:2014,Uhlemann:2015,Okumura:2015}).
However, our galaxy sample lies at a relatively high redshift $z\sim 1.3$ where density perturbations have not grown as much as today, we will use only the clustering information on large scales, so the nonlinear effect is weaker. 
Also the error on the measurements of the correlation function for the FastSound sample is much larger than the improvement of the accuracy of the models.
Considering these facts, we thus do not consider these nonlinear RSD models and only use the 5 models mentioned above, equations (\ref{eq:kaiser}), (\ref{eq:scoccimarro}) and (\ref{eq:nl_kaiser}). 

\subsection{Tests against mocks}\label{sec:mock_test}

In order to see whether these 5 theoretical models can provide unbiased constraints on the growth rate parameter, 
we analyze the mock galaxy catalogs and see if the input cosmological parameters for the simulation are reproduced.  
Here, as the observed correlation function we consider the average of the correlation functions measured from the 640 mock catalogs, and use the covariance matrix for the correlation function measured for the FastSound galaxy sample estimated by equations (\ref{eq:cov}) and (\ref{eq:cov2}).

In the upper panel of figure \ref{fig:mock_fsig8}, we show the constraints $f\sigma_8$ from our mock catalogs as a function of the minimum separation $r_{\rm min}$. 
The maximum separation $r_{\rm max}$ is fixed to $80\himpc$.
The red points are the result analyzed using the linear RSD model [equation (\ref{eq:kaiser})], and the other points are obtained using the 4 nonlinear models (eqs. \ref{eq:nl_kaiser} and \ref{eq:scoccimarro}) with $\sigma_{\rm v}$ fixed to zero or being a free parameter. 
The bias parameter $b\sigma_8$ and the velocity dispersion parameter $\sigma_{\rm v}$ for $G\neq 1$ models are marginalized over. 
For all the minimum separation values $r_{\rm min}$ probed, the results analyzed with the 5 models all reproduce the input parameter of the simulation that is denoted as the horizontal dashed line within the $1-\sigma$ error. 
The best fitting parameter of $f\sigma_8$ analyzed with linear theory is slightly higher than the true value, consistent with the high-$z$ analysis of RSD by \citet{Marulli:2015}.
As we increase the minimum scale, the constrained $f\sigma_8$ approaches the input value, which also explains the behaviors of the multipoles at high redshift shown by \citet{Okumura:2012b}.
Even though we let $\sigma_{\rm v}$ be a free parameter, the constraints on $\sigma_{\rm v}$ are consistent with zero as shown in the upper panel of figure \ref{fig:mock_sigv}, but the best fitting parameter $f\sigma_8$ is biased toward higher values. 
This degeneracy can be well illustrated by the joint constraint on $f\sigma_8$ and $\sigma_{\rm v}$ in the lower panel of figure \ref{fig:mock_sigv}.
The positive correlation between the two parameters comes from the fact that increasing $f\sigma_8$ enhances the clustering amplitude coherently while increasing $\sigma_{\rm v}$ suppresses the small-scale clustering. 

Note that because we define the fitting scale range by the three dimensional separation $r$, not by the transverse separation $r_p$, the data below $r_{\rm min}$ along the line of sight comes into the cosmological constraints due to the nonlinear velocity dispersion and it may bias the constraints. However, our galaxy sample is so sparse that the correlation function at scales less than $10\himpc$ does not contain much cosmological information as seen in figure \ref{fig:mock_fsig8}, thus we do not expect it to be the case for our analysis. The shift of the constraint on $f\sigma_8$ by allowing $\sigma_{\rm v}$ to be free is indeed small, consistent with the result at similar redshift by \citet{Marulli:2015} who adopted the cut in the $(r_p,r_\pi)$ plane.

\begin{figure}[t]\begin{center}
\FigureFile(70mm,60mm){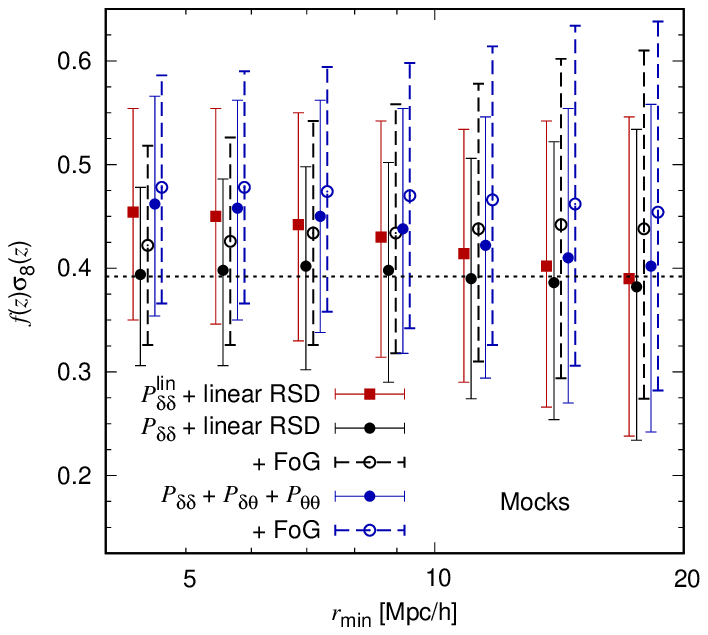}
\FigureFile(70mm,60mm){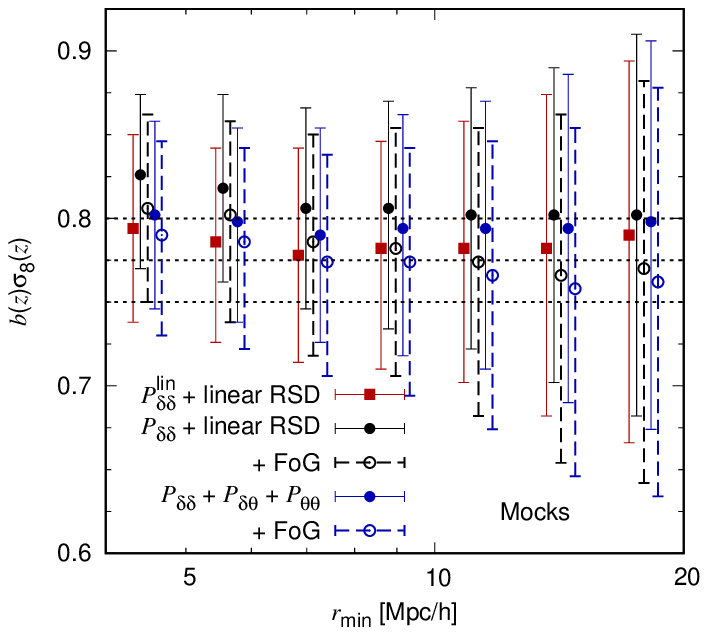}
  \end{center}
 \caption{Constraints on $f\sigma_8$ ({\it upper panel}) and $b\sigma_8$ ({\it lower panel}) 
 as a function of the minimum separation $r_{\rm min}$ for mock galaxy samples. 
The maximum separation is fixed to $r_{\rm max}=80\himpc$. The constraints are obtained using 5 RSD models: 
 linear Kaiser model ({\it red}), the nonlinear matter power spectrum with Kaiser RSD with $\sigma_{\rm v}=0$ ({\it black solid}) and $\sigma_{\rm v}$ being a free parameter ({\it black dashed}), and nonlinear Kaiser terms of $P_{\delta\delta}$, $P_{\delta\theta}$, $P_{\theta\theta}$ with $\sigma_{\rm v}=0$ ({\it blue solid}) and $\sigma_{\rm v}$ being a free parameter ({\it blue dashed}).
The horizontal line in the upper panel represents the linear theory prediction computed from the input cosmological parameters for the simulation.  
The three horizontal lines in the lower panel are added to help to see deviation of the $b\sigma_8$ values from constants. 
}
\label{fig:mock_fsig8}
\end{figure}

\begin{figure}[t]\begin{center}
\FigureFile(70mm,70mm){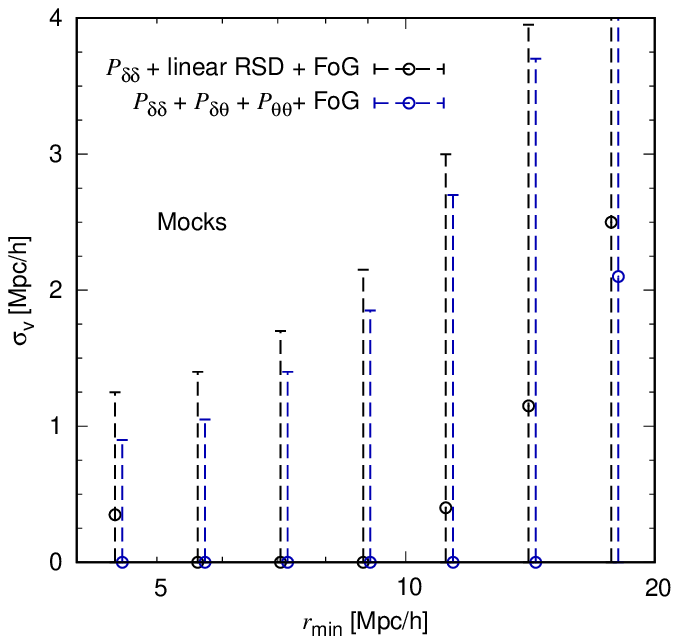}
\FigureFile(70mm,70mm){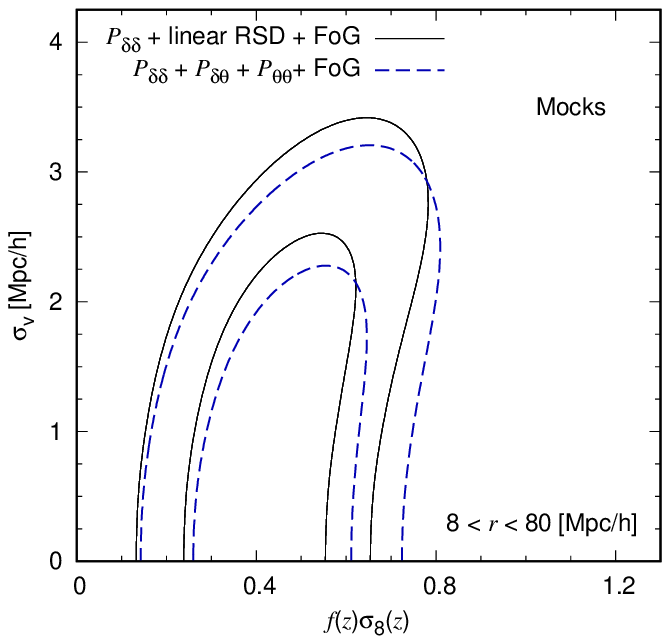}
  \end{center}
 \caption{({\it Upper panel}) Same as figure \ref{fig:mock_fsig8} but a constraint on $\sigma_{\rm v}$ for the two FoG models ($\sigma_{\rm v}\neq 0$): 
the nonlinear matter power spectrum with Kaiser RSD ({\it black dashed}) and nonlinear Kaiser terms of $P_{\delta\delta}$, $P_{\delta\theta}$, $P_{\theta\theta}$ ({\it blue dashed}).
({\it Lower panel}) Joint constraint on $f(z)\sigma_8(z)$ and $\sigma_{\rm v}$ for the two models. The inner and outer contours respectively show the $1-\sigma$ and $2-\sigma$ confidence levels. 
}
\label{fig:mock_sigv}
\end{figure}

The best RSD model among the five for our mock sample is the linear Kaiser model with the nonlinear matter power spectrum $P_{\delta\delta}$ [equation (\ref{eq:nl_kaiser})], as shown as the black filled-symbol points. It provides an unbiased constraint on $f\sigma_8$ even when we adopt very small values of $r_{\rm min}$. 
However, we cannot conclude that the RSD model with $\sigma_{\rm v}=0$ is the best model to analyze the real FastSound galaxy data since our mock data might be too simplistic. We adopted the central HOD with zero relative velocity to the halo center. Such a model neglects the substantial relative velocity found for the brightest halo galaxies by \citet{Skibba:2011} (see also \cite{Hikage:2012} for the off-centering effect of central galaxies), although emission line galaxies reside in less massive halos so that such an effect is expected to be smaller.
Thus in the next section, while we adopt the linear Kaiser model with the nonlinear $P_{\delta\delta}$ and no velocity dispersion to obtain main cosmological results from the FastSound data, we will examine the case when $\sigma_{\rm v}$ is treated as a free parameter and see if the constraint on $f\sigma_8$ is biased. 

Although the model reproduces the input $f\sigma_8$ of the mocks even when the small scale data are used, $r_{\rm min}$ value, $\sim 5\himpc$, we will perform a conservative analysis and choose $r_{\rm min}$ where both this model and the linear RSD model give the best fitting $f\sigma_8$ consistent with the true value within $10\%$, which is $\sim 27\%$ percent of our statistical error. 
We thus use the measurements of the correlation function only at large scales, choosing $r_{\rm min}=8\himpc$ as the default minimum scale.
The assumption of $\sigma_{\rm v}$ fixed to zero for analyzing the FastSound galaxy sample will be a reasonable assumption because of the result in the upper panel of figure \ref{fig:mock_sigv} and the fact that the shapes of the multipole correlation functions for the data and the mocks are exactly the same at small scales where the effect of the nonlinear velocity dispersion is largest, down to $r_{\rm min}$. 
A detailed study focusing on the small scale clustering will be performed in future paper (\hikage).

Next, we test the assumption of the linear bias made in our modeling. 
The lower panel of figure \ref{fig:mock_fsig8} is the same as the upper panel but shows the best fitting value of $b\sigma_8$ as a function of the minimum separation $r_{\rm min}$.
The three horizontal lines are added to help to see deviations of the $b\sigma_8$ values from constants at low $r_{\rm min}$. 
For each of the five models the difference of the best fitting parameter of the bias $b\sigma_8$ is $\sim 5\%$ between $r_{\rm min}=4\himpc$ and $16\himpc$.
Particularly, the model of the nonlinear density power spectrum of dark matter with the linear Kaiser and no velocity dispersion has a constant bias between $r_{\rm min}=8\himpc$ and the larger $r_{\rm min}$. We thus conclude that the simple linear, constant bias is a reasonable model for our sample.

\subsection{Redshift blunder correction}\label{sec:blunder}
To constrain cosmological parameters from our FastSound sample, we also need to model the effect of two types of redshift blunders, noise lines and OIII doublets, 
as we have mentioned in section \ref{sec:fastsound}. 
They smear the anisotropies in the correlation function. 
Without any assumption, the measured correlation function can be decomposed into three terms: the auto correlation of real galaxies, that of the auto correlation of fake galaxies (the blunders), and their cross correlation (see e.g., \cite{Okumura:2015} for a decomposition scheme of a measured redshift-space power spectrum). 
The simplest approximation is that the fake signals are assumed to be distributed randomly so that the latter two terms are equal to zero, and the true clustering amplitude is obtained by the simple scaling of 
$(1-f_{\rm blund})^2 \equiv (1-f_{\rm fake})^2(1-f_{\rm OIII})^2$ \citep{Blake:2010}.
This approximation may be too specific, but if the fraction $f_{\rm blund}$ is small, e.g., $f_{\rm blund}=0.1$, the cross correlation between H$\alpha$ and fake galaxies is suppressed by 0.18 and the auto correlation of the fake galaxies by 0.01, the approximation will work well because the correlation of the fake galaxies itself is much weaker than that of H$\alpha$ galaxies. 
We mainly analyze the sample with line $S/N>4.5$ which corresponds to $f_{\rm blund}=0.071$ as we have seen in section \ref{sec:fastsound}, but we will test this approximation by investigating how much the final constraints on the growth rate are shifted by changing the fraction to $f_{\rm blund}=0.124$ ($S/N>4.0$) and $f_{\rm blund}=0.054$ ($S/N>5.0$).

\section{Analysis and results}\label{sec:analysis}

\subsection{Setup for parameter fits}
We present constraints on cosmological models by comparing the observed monopole and quadrupole with the corresponding theoretical models. 
We adopt a simple $\chi^2$ statistics to give constraints on cosmological models.
Let $N_{\rm bin}$ be the number of bins for statistics used to obtain cosmological constraints $r_{\rm min}<r<r_{\rm max}$, then
the $\chi^2$ statistics is given by
\be
\chi^2({\bf \theta})=\sum^{2N_{\rm bin}}_{i=1}\sum^{2N_{\rm bin}}_{j=1} \Delta_i C_{ij}^{-1}\Delta_j,
\ee
where 
$\Delta_i = \xi^{s,obs}_\ell(r_i)-\xi^{s,th}_\ell(r_i;\theta)$ is the difference between the observed correlation function and theoretical prediction with ${\bf \theta}$ being a parameter set to be constrained, and $C^{-1}$ is the inverse of 
the covariance matrix we obtained in section \ref{sec:covariance}. The factor of $2$ in front of $N_{\rm bin}$ comes from
the fact that we use the monopole ($\ell=0$) and quadrupole ($\ell=2$) for the analysis.
The likelihood function $L$ is proportional to $\exp{(-\chi^2/2) }$. Then the $1\sigma$ confidence level (CL) interval 
is determined by the region where the integration of $L$ over a given cosmological parameter becomes $68\%$ of the entire parameter space.

As we discussed in the previous section, we adopt the model of the linear Kaiser with the nonlinear matter power spectrum, with the growth rate and bias being free parameters, $\theta=(f\sigma_8,b\sigma_8)$. 
The nonlinear velocity dispersion parameter $\sigma_{\rm v}^2$ is fixed to zero to obtain the main constraint because the best fitting value of $\sigma_{\rm v}$ against mocks is zero and the shapes of the correlation functions at small scales for the data and mocks were the same. We performed the $\chi^2$ analysis for the data sample by varying $\sigma_{\rm v}$ and found that the value which gave the minimum $\chi^2$ was indeed $\sigma_{\rm v}=0$, as we will show below. 
Another parameter, the fraction of redshift blunders, is fixed by the observation as $f_{\rm blund}=0.071$ for our FastSound galaxy sample with the line threshold $S/N>4.5$. 

\subsection{Constraints on linear growth rate}

We have investigated at which scales we can safely use the clustering information using the mock catalogs in section \ref{sec:mock_test}, 
and chose a minimum separation $r_{\rm min}=8\himpc$. We fix the maximum separation $r_{\rm max}=80\himpc$, but changing $r_{\rm max}$ does not 
change the cosmological constraints because the data at $r>40\himpc$ are noisy. 

In the top panel of figure \ref{fig:fsig8_bsig8} we present the joint constraints on the linear growth rate and the galaxy bias, $(f\sigma_8,b\sigma_8)$ from the correlation function of the FastSound sample at $8<r<80\himpc$. The parameter set with the minimum value of $\chi^2$ is $(f\sigma_8,b\sigma_8)=(0.478,0.818)$, shown as the filled triangle, where $\chi^2_{\rm min}=10.21$ and the degrees of freedom are $2\times 10-2=18$.
For comparison, we also show the result when we adopt $r_{\rm min}=16 \himpc$. 
There is a strong degeneracy, $bf\sim $ constant, particularly if only the data at large scales are used because the constraints mostly come from the 
amplitude of the redshift-space correlation function. Since the linear RSD constrains a combination $\beta=fb^{-1}$, the direction of degeneracy changes 
by including smaller scale data. 
For comparison, the constraints from our mock catalogs, analyzed in section \ref{sec:mock_test}, are plotted in the same panel with the red dashed contours. 
The parameters with the minimum $\chi^2$ are $(f\sigma_8,b\sigma_8)=(0.396,0.806)$, so the FastSound data give systematically higher values. 
This may imply that our model of the redshift blunders is too simple; multiplying the correlation function by the factor $(1-f_{\rm blund})^{-2}$ enhances the amplitude, namely $b\sigma_8$, while keeping $\beta=fb^{-1}$ as a constant thereby $f\sigma_8$ is also enhanced. Nevertheless the constraints from the data and mocks are consistent with each other at $1-\sigma$.

\begin{figure}[t]\begin{center}\FigureFile(75mm,75mm){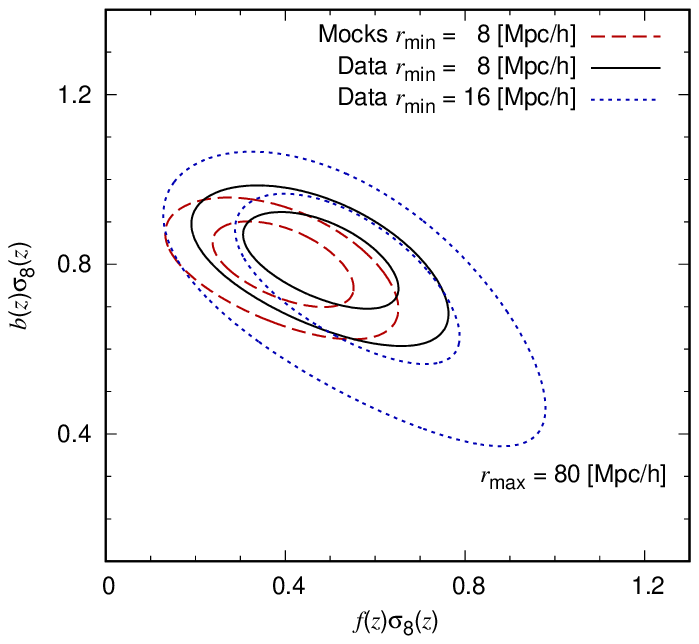}
\FigureFile(80mm,80mm){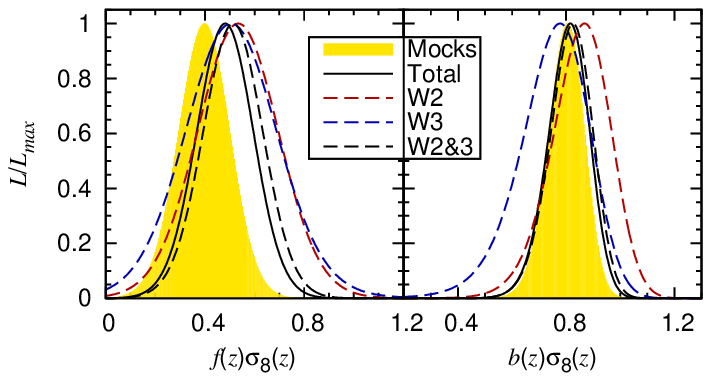}
  \end{center}
 \caption{({\it Top}) Joint constraint on $f(z)\sigma_8(z)$ and $b(z)\sigma_8(z)$. The thick, black solid contours show the $1-\sigma$ and $2-\sigma$ confidence levels from inside to outside from the data, while the dashed red contours are from the mock catalogs. 
For the analysis the correlation function at $8<r<80\himpc$ is used. 
The velocity dispersion $\sigma_{\rm v}$ is fixed to zero and the parameter set with the minimum $\chi^2$ is denoted as the filled triangle. The result with the minimum $\chi^2$ when $\sigma_{\rm v}$ is marginalized over is shown as the open triangle.
For comparison, the result from the data but with $r_{\rm min}=16\himpc$ is presented as the blue dotted contours. 
({\it Bottom}) 1-d likelihood functions of the growth rate $f\sigma_8$ and $b\sigma_8$ from each of the W2 (red dashed) and W3(blue dashed), and of the combination (W2+W3). 
 The black solid line is from the whole sample. 
 The functions colored by yellow are the result from the mock catalogs.
 }
\label{fig:fsig8_bsig8}
\end{figure}

\begin{figure}\begin{center}\FigureFile(75mm,75mm){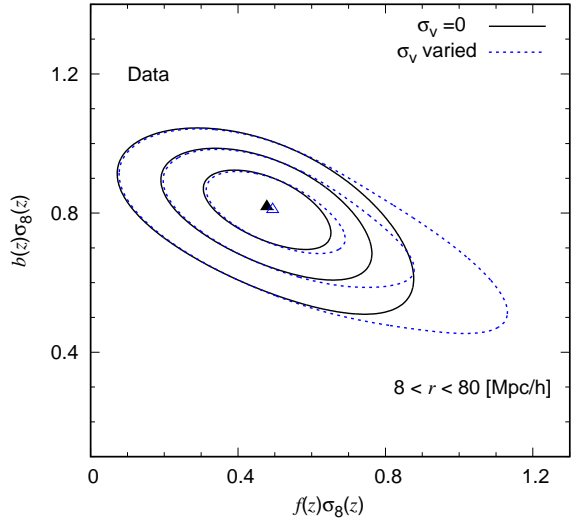}
  \end{center}
 \caption{Joint constraint on $f(z)\sigma_8(z)$ and $b(z)\sigma_8(z)$ when the velocity dispersion is fixed to $\sigma_{\rm v}=0$ ({\it solid contours}) and is varied and marginalized over ({\it dashed contours}). The contours show the 1, 2 and $3-\sigma$ confidence levels from inside to outside. 
The parameter set for the maximum likelihood is respectively shown as the filled and open triangles. 
 }
\label{fig:fsig8_bsig8_sigv}
\end{figure}

The bottom panels of figure \ref{fig:fsig8_bsig8} present the 1-d likelihood functions with another parameter marginalized over as 
$L(\theta_i) \propto \int d\theta_j \exp{\left[-\frac{1}{2}\chi^2(\theta_i,\theta_j)\right] }$ where ${\boldsymbol \theta}=(f\sigma_8,b\sigma_8)$. 
The results are normalized by the maximum likelihood, $L(\theta_i)/L_{\rm max}$. 
The best fitting parameter for the growth rate parameter is $f\sigma_8=0.482^{+0.116}_{-0.116}$ for the fits to $8<r<80\himpc$ ($1\sigma$ CL),  
which excludes the isotropic clustering (with no RSD) with $4.1\sigma$. 
It is also consistent with a prediction from $\Lambda$CDM model with general relativity $f\sigma_8\sim 0.392$.
The bias parameter for the FastSound galaxy sample is constrained to $b\sigma_8 = 0.814^{+0.076}_{-0.080}$. 
If we assume the value of $\sigma_8$ predicted by the recent CMB experiments, the bias of our FastSound galaxy sample is $b\sim 1.9$, consistent with the HOD analysis of the same galaxy sample by \hikaget. The host halo mass obtained by the HOD analysis and the constrained bias are also consistent with the clustering analysis of from the Hi-Z Emission Line Survey (HiZELS, \cite{Geach:2012}) which used the H$\alpha$ emitter sample with the similar flux limit at similar redshift to our sample. 
The relationship between the FastSound galaxies and the underlying dark matter density will be investigated in more detail in \hikaget.
The 1-d constraint on the bias parameter from the mock catalogs, colored by yellow at the bottom right panel of figure \ref{fig:fsig8_bsig8} is in good agreement with the results from the data. 

We perform the same analysis allowing $\sigma_{\rm v}$ to vary and compute $\chi^2(\theta)$ with $\theta=(f\sigma_8,b\sigma_8,\sigma_{\rm v})$. For fixed values of $f\sigma_8$ and $b\sigma_8$, $\sigma_{\rm v}=0$ gives the minimum $\chi^2$. We show the likelihood contours of $(f\sigma_8,b\sigma_8)$ after marginalizing over the $\sigma_{\rm v}$ parameter and compare to the result with $\sigma_{\rm v}=0$ in figure \ref{fig:fsig8_bsig8_sigv}. The parameter set for the maximum likelihood is $(f\sigma_8,b\sigma_8)=(0.494,0.810)$, denoted as the open triangle in figure \ref{fig:fsig8_bsig8_sigv}, very close to the result with $\sigma_{\rm v}=0$ (the filled triangle). When the bias parameter $b\sigma_8$ is also marginalized over, our constraint on the growth rate is $f\sigma_8=0.494^{+0.126}_{-0.120}$ ($1-\sigma$). Thus although the error bound in the growth rate becomes larger by $6\%$ the best fitting value is almost unchanged ($\sim 2.5\%$ shift). 

The bottom panels of figure \ref{fig:fsig8_bsig8} also show 1d constraints on the growth rate and bias parameters 
for each of the main two CFHTLS survey fields, W2 and W3, shown as the red and blue dashed curves, respectively.
The constraints from the combination of the two fields are shown as the black dashed curve. 
The other two fields, W1 and W4 are so small that adding the data from these two fields makes an marginal improvement to the final constraints. 
However, each field gives almost the same best fitting parameters. 

The best fitting model for the anisotropic correlation function $\xi^s(r_p,r_\pi)$ is computed using equation (\ref{eq:xi_multi}) and shown as the line contours 
in figure \ref{fig:2dxi}, those for the multipoles $\xi^s_0$ and $\xi^s_2$ are shown as the black solid lines in figures 
\ref{fig:mono} and \ref{fig:quad}, respectively. The model correlation functions are multiplied by $(1-f_{\rm blund})^{-2}$. 

To see the validity of the simple model of redshift blunders considered in section \ref{sec:blunder}, we repeat the same analysis by changing the threshold of the line signal-to-ratio from the default value of $S/N>4.5$. When we lower the threshold to $S/N>3.0$, the real emission lines are only less than the half of the detected lines (see Table \ref{tab:gal}) and the model we considered is too simplistic. Indeed the clustering anisotropy is smeared out by using the sample and $f\sigma_8$ is poorly constrained, consistent with zero detection of RSD. 
On the other hand, the best fitting value of $f\sigma_8$ has been shifted only by $5\sim 6\%$ when changing the threshold to $S/N>4.0$ or $>5.0$, which is much smaller than the $1-\sigma $ error of the constraint  ($\sim 25\%$). 
We thus conclude that the line threshold we adopt is strict enough that our model for the redshift blunders is a reasonable assumption. 

\subsection{Alcock-Paczynski effect}
The distance to each galaxy is determined from the redshift by assuming fiducial cosmology, $\Omega_m=1-\Omega_\Lambda=0.27$ in our case. 
If the assumed parameter is incorrect, geometric anisotropy is induced to the clustering pattern, known as the Alcock-Paczynski (AP) effect \citep{Alcock:1979}.
The AP distortion alters the radial and transverse distances, thus are sensitive to the Hubble parameter $H(z)$ and the angular-diameter distance $D_A(z)$ and 
known to degenerate with $f(z)$ from the RSD anisotropy \citep{Ballinger:1996,Matsubara:1996,Seo:2003,Blazek:2014}. 

To see the effect of the AP effect on our constraint of $f\sigma_8$, we perform a simple test which was done by \citet{Contreras:2013} for the correlation analysis of the WiggleZ data. We vary the value of $\Omega_m$ for converting $z$ to the comoving distance, and measure the correlation function for each $\Omega_m$. We then analyze it using the matter power spectrum $P_{\delta\delta}$ with the same value of $\Omega_m$ and the other cosmological parameters being fixed to the fiducial values, and constrain the set of RSD parameters $(f\sigma_8, b\sigma_8)$. The resulting $f\sigma_8$ as a function of $\Omega_m$ after marginalizing over $b\sigma_8$ is shown in figure \ref{fig:fsigma8_ap}.
Around the values of $\Omega_m$ of interest, $0.2<\Omega_m<0.4$, $f\sigma_8$ is nearly constant, thus our constraint is unlikely to shift due to the AP effect. A similar conclusion was reached for the lower-$z$ but larger-volume, VIPERS survey \citep{de-la-Torre:2013}. 
However, marginalizing over the AP effect makes the $f\sigma_8$ constraint looser, as nicely demonstrated by \citet{Ruiz:2015} (it could be about a factor of two at the redshift of the FastSound survey).

For comparison, we also show the theoretical curve of $f\sigma_8$ with the cosmological parameters fixed to the $\Lambda$CDM model but varying $\Omega_m$. 
Note that the clustering amplitude is fixed at $z=0$ to $\sigma_8=0.82$, thus the amplitude of the primordial scalar fluctuation $A_s$ is different for different $\Omega_m$ values.

\begin{figure}\begin{center}\FigureFile(75mm,75mm){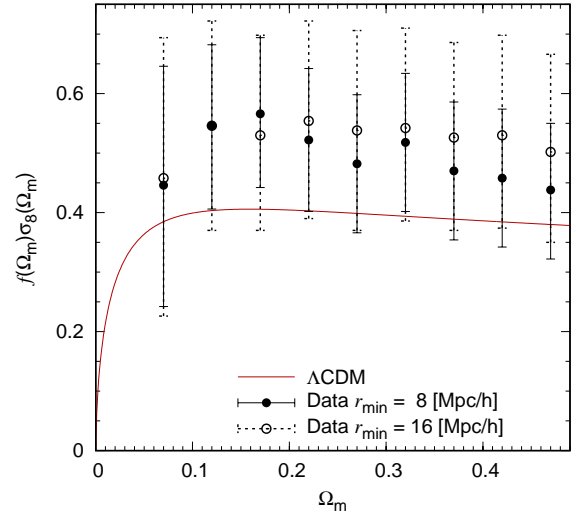}
  \end{center}
 \caption{Constraints on $f\sigma_8$ as a function of $\Omega_m$ for converting $z$ to the distance and for analyzing the data. 
 The filled-symbol points with the solid error bars and the open-symbol points with the dotted error bars are respectively the results when the data are 
 analyzed for $r_{\rm min}=8 \himpc$ and $16\himpc$. The red line is the prediction for $f\sigma_8$ as a function of $\Omega_m$ and with other parameters being 
 fixed to the fiducial $\Lambda$CDM values. Note that the amplitude of the perturbation in the theoretical curve is fixed at $z=0$ to $\sigma_8=0.82$.}
\label{fig:fsigma8_ap}
\end{figure}

\section{Gravity theories as a function of redshift}\label{sec:fsigma8_z}

In this section, we present a demonstration what kind of insights we can obtain for gravity theory models by combining the constraint on $f\sigma_8$ from the FastSound survey with those from low-$z$ RSD surveys. 
In section \ref{sec:fsigma8_rsd} we first present the constraints on $f\sigma_8$ from various RSD observations as a function of redshift to determine the normalization of the density fluctuation, $\sigma_8$, using only RSD.
Then we compare the observed constraints to various modified gravity models in section \ref{sec:fsigma8_mg}. 
Note, however, that we do not intend to test which model is preferred by our observations, but 
rather simply demonstrate that our $z\sim 1.4$ measurement can be used as a high-$z$ anchor for $f\sigma_8$, independent of the CMB experiments. 
In section \ref{sec:fsigma8_cmb}, we test the consistency of the $f\sigma_8$ constraint from the FastSound survey with the CMB anisotropy probes. 

\subsection{Growth rate from FastSound and lower-$z$ surveys}\label{sec:fsigma8_rsd}

Models of gravity theories have never been tested at redshift $1<z<2$, thus our RSD analysis of the FastSound galaxy sample provides the first test in the redshift range, $1.19<z<1.55$. 
Figures \ref{fig:fsigma8_table_lss} and \ref{fig:fsigma8_table_cmb} show the constraints on $f(z)\sigma_8(z)$ as a function of $z$ obtained from our FastSound sample at $1.19<z<1.55$ together with the previous studies at lower redshifts at $z<1$. 
These constraints include the studies of the 6dFGS \citep{Beutler:2012}, the SDSS main galaxies \citep{Howlett:2015}, the 2dFGRS \citep{Song:2009a}, the SDSS LRG \citep{Samushia:2012}, the BOSS LOWZ \citep{Chuang:2013}, the BOSS CMASS \citep{Reid:2012}, WiggleZ \citep{Blake:2011}, VVDS \citep{Guzzo:2008} and VIPERS \citep{de-la-Torre:2013} surveys.

Since not all the data points are from independent survey regions (redshift and angular positions), the covariance between the data points needs to be taken into account for the accurate analysis. 
However, we do not intend to present detailed tests for modified gravity theories, but we rather want to demonstrate the importance of RSD analysis over wide redshift ranges. We thus simply pick up the tightest constraint if two survey areas are largely overlapped. We use only the data points denoted as the filled squares as well as our constraint from the FastSound survey in figure \ref{fig:fsigma8_table_lss}. They are, 6dFGS with the effective redshift $z_{\rm eff}=0.067$, the two points from SDSS LRG with $z_{\rm eff}=0.25$ and $z_{\rm eff}=0.37$, the SDSS CMASS with $z_{\rm eff}=0.57$, the highest redshift bin out of the four from the WiggleZ with $z_{\rm eff}=0.78$, the VIPERS with $z_{\rm eff}=0.8$. With this choice, all the data points are uncorrelated except for the $2.1\%$ correlation between the CMASS and the higher-redshift bin of the LRG (see \cite{Alam:2016}). 
Using the 7 data points of $f\sigma_8$, we compute the $\chi^2$ for theoretical predictions of gravity theories including GR with the amplitude of $f\sigma_8$ being a free parameter. 
The $\Lambda$CDM model plus GR with the best fitting amplitude is shown as the solid line in figure \ref{fig:fsigma8_table_lss}. 

\begin{figure*}\begin{center}\FigureFile(160mm,160mm){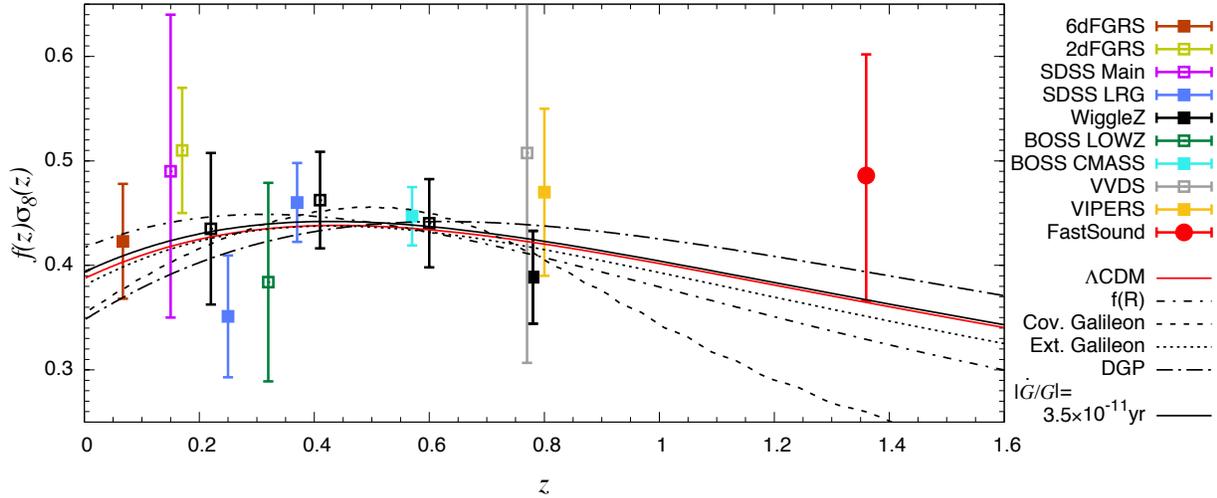}
  \end{center}
 \caption{Constraints on the growth rate $f(z)\sigma_8(z)$ as a function of redshift at $0<z<1.55$. The constraint obtained from our FastSound sample at $1.19<z<1.55$ is plotted as the big red point.
 The previous results include the 6dFGS, 2dFGRS, SDSS main galaxies, SDSS LRG, BOSS LOWZ , WiggleZ, BOSS CMASS, VVDS, and VIPERS surveys at $z<1$. 
 A theoretical prediction for $f\sigma_8$ from $\Lambda$CDM and general relativity with the amplitude determined by minimizing $\chi^2$ is shown as the red solid line. 
 The data points used for the $\chi^2$ minimization are denoted as the filled-symbol points while those which are not used are denoted as the open-symbol points. 
 The predictions for $f\sigma_8$ from modified gravity theories with the amplitude determined in the same way are shown as the thin lines with different line types; 
 $f(R)$ gravity model (dot-short-dashed), the covariant Galileon model (dashed), the extended Galileon model (dotted), DGP model (dot-dashed), and the early, time varying gravitational constant model (black solid).
}
\label{fig:fsigma8_table_lss}
\end{figure*}

\subsection{Modified gravity models}\label{sec:fsigma8_mg}

On the scales probed by large-scale structure surveys, 
the growth rate $f$ generally obeys a simple evolution 
equation \citep{Baker:2014,Leonard:2015}:
\begin{eqnarray}
& f^\prime+q(x)\,f+f^2=\frac{3}{2}\Omega_m 
\xi \label{f_eq},\\
& \mathrm{where}\quad\quad q(x)=\frac{1}{2}
\left[1-3\,w(x) 
(1-\Omega_m(x)) \right]\,, \label{qdef}
\end{eqnarray}
here, $x=\ln a$ (where $a$ is the scale factor) and prime denotes derivatives with regards to $x$. 
The effective equation of state for the non-matter sector is 
denoted by $w(x)$ and $\xi\equiv\mu/\gamma$ encodes deviations from GR/Newtonian gravity: 
$\mu=G_{\rm eff}/G_0$ where $G_{\rm eff}$ is the effective Newton's constant and $G_0$ is the `bare' constant that would appear in the action. The gravitational slip parameter is 
defined by $\gamma=\Phi/\Psi$, where $\Psi$ and 
$\Phi$ are gravitational potentials appearing in the 
perturbed metric $ds^2=-(1+2\Psi)dt^2+a^2(t) (1-2\Phi)\delta_{ij}dx^idx^j$.
Finally, we can obtain $\sigma_8$ by integrating $f$ with the appropriate boundary conditions
(such as, for example, the normalization arising from the CMB). One can immediately see that $f$ is sensitive to both the expansion rate (and the corresponding equation of state, $w(x)$) and gravitational physics (via $\xi$). Typically models that attempt to explain the accelerated expansion through modifications of gravity affect both aspects: expansion and growth \citep{Clifton:2012}.

In figure \ref{fig:fsigma8_table_lss} we plot a number of alternative models that we now discuss in turn in the following subsections. 
As we did for the $\Lambda$CDM model with GR, for each gravity model the amplitude of $f\sigma_8$ is fitted against the 7 data points and the model with the best fitting amplitude is presented in the figure. We do not intend to perform a test of each model.

\subsubsection{$f(R)$ gravity}
To begin with we consider models in which the Einstein-Hilbert action is modified as follows
\begin{eqnarray}
\frac{1}{16\pi G_0} 
\int d^4 x\sqrt{-g}R\rightarrow 
\frac{1}{16\pi G_0} \int d^4 x \sqrt{-g}f(R)\,,
\end{eqnarray}
where $f(R)$ is constructed to mimic, as closely as possible, the observed accelerated expansion.
A notable example is given by \citet{Hu:2007}
\be
f(R)=R-\lambda R_c \frac{(R/R_c)^{2n}}{(R/R_c)^{2n}+1},
\label{fRmodel}
\ee
where $n$, $\lambda$ and $R_c$ are positive constants. 
Similar viable $f(R)$ models have been independently proposed by 
\citet{Starobinsky:2007}, \citet{Appleby:2007} and \citet{Tsujikawa:2008}.

For $n$ and $\lambda$ of the order of 1 we have that 
$R_c \approx H_0^2$, where $H_0$ is  
the today's Hubble constant. 
In the high-curvature regime characterized by $R \gg R_c$, 
the model (\ref{fRmodel}) reduces to 
$f(R) \simeq R-\lambda R_c[1-(R/R_c)^{-2n}]$, so it is close 
to the $\Lambda$CDM model. 
As $R$ decrease to the order of $R_c$,  
the deviation from the $\Lambda$CDM model arises 
at low redshifts. This is the regime in which the modification 
of gravity manifests itself in the observations of RSD.
Provided that the stability condition 
$0<Rf_{,RR}/f_{,R} \le 1$ (where $f_{,R}=df/dR$) is satisfied, the solution finally approaches a de Sitter 
solution characterized 
by $Rf_{,R}=2f$ (\cite{Amendola:2007}).
In this case the  the effective 
gravitational coupling in $f(R)$ gravity is 
given by \citep{Tsujikawa:2007,de-Felice:2011}
\begin{equation}
G_{\rm eff}=\frac{G_0}{f_{,R}}
\frac{1+4r/3}{1+r}\,,\qquad 
r=\left( \frac{k}{am_{\phi}} \right)^2\,.
\end{equation}
where  $m^2_\phi\simeq f_{,R}/(3f_{,RR})$
and we have that the  
$f(R)$ model (\ref{fRmodel}) exhibits the 
gravitational interaction stronger than that in  
the $\Lambda$CDM model at low redshifts.

As an example, we choose $n=2$ and $\lambda=2$ and compute the $\chi^2$ statistics by changing the normalization of $f\sigma_8$ as we have done for GR above. The resulting $f\sigma_8$ as a function of $z$ with the best fitting amplitude at the scale $k^{-1}=30\himpc$ is shown as the dot-dashed line in figure \ref{fig:fsigma8_table_lss}.
Because the $f(R)$ gravity model exhibits stronger gravity than GR, fitting the $f(R)$ model to the RSD measurements gives $f\sigma_8$ smaller than the $\Lambda$CDM model at higher redshift. 

\subsubsection{Dvali-Gabadadze-Porrati braneworld}

An alternative model we consider is the 
Dvali-Gabadadze-Porrati (DGP) braneworld \citep{Dvali:2000}, 
in which a 3-brane is embedded in a 5-dimensional (5D) Minkowski 
bulk spacetime with an infinitely large extra dimension.
In the effective 4-dimensional (4D) picture, the Friedmann 
equation on the flat FLRW brane is given by $H^2-\epsilon H/r_c=
\kappa_{4}^2\rho_m/3$, where $\epsilon=\pm 1$
and $r_c=\kappa_{(5)}^2/(2\kappa_{(4)}^2)$ is 
a length scale determined by the ratio of 5D and 4D 
gravitational constants $\kappa_{(5)}$ and $\kappa_{(4)}$.
For the branch $\epsilon=+1$, there is a de Sitter solution 
characterized by the Hubble parameter $H_{\rm dS}=1/r_c$.
We include this model because it realizes (as we shall see) 
$G_{\rm eff}<G$; unfortunately it is
associated with the existence of ghosts  (\cite{Nicolis:2004}). 

On the scale of surveys we have that the
effective Newton's constant satisfies (\cite{Lue:2004}, \cite{Koyama:2006}):
\begin{equation}
G_{\rm eff}=\left[ 1+\frac{1}{3\beta(t)} \right]G_0\,,\qquad
\beta(t) \equiv 1-2Hr_c \left( 1+\frac{\dot{H}}{3H^2} \right).
\end{equation}
Since $Hr_c \gg 1$ and $\dot{H}/H^2 \simeq -3/2$ in the 
deep matter era, it follows that $|\beta| \gg 1$ and hence 
$G_{\rm eff} \simeq G$. As the background trajectory 
approaches the de Sitter solution characterized by 
$Hr_c=1$ and $\dot{H}=0$, we have that $\beta=-1$ 
and $G_{\rm eff}=2G/3$. The DGP model gives rise to 
weaker gravity due to the gravitational leakage to the 
extra dimension.  

Since the DGP model predicts a weaker gravitational interaction 
on cosmological scales, fitting the amplitude of 
$f(z)\sigma_8(z)$ to RSD measurements without using the bound 
of $\sigma_8(0)$ from CMB measurements gives rise to 
$f(z)\sigma_8(z)$ larger than that of the $\Lambda$CDM model 
at high redshifts ($z>1$). The best-fit curve of the DGP model 
is plotted as the dot-long-dashed line in 
Fig.~\ref{fig:fsigma8_table_lss}, which exhibits a notable 
deviation from the $\Lambda$CDM model and $f(R)$ gravity 
at the redshift associated with the FastSound measurement.


\begin{figure*}\begin{center}\FigureFile(160mm,160mm){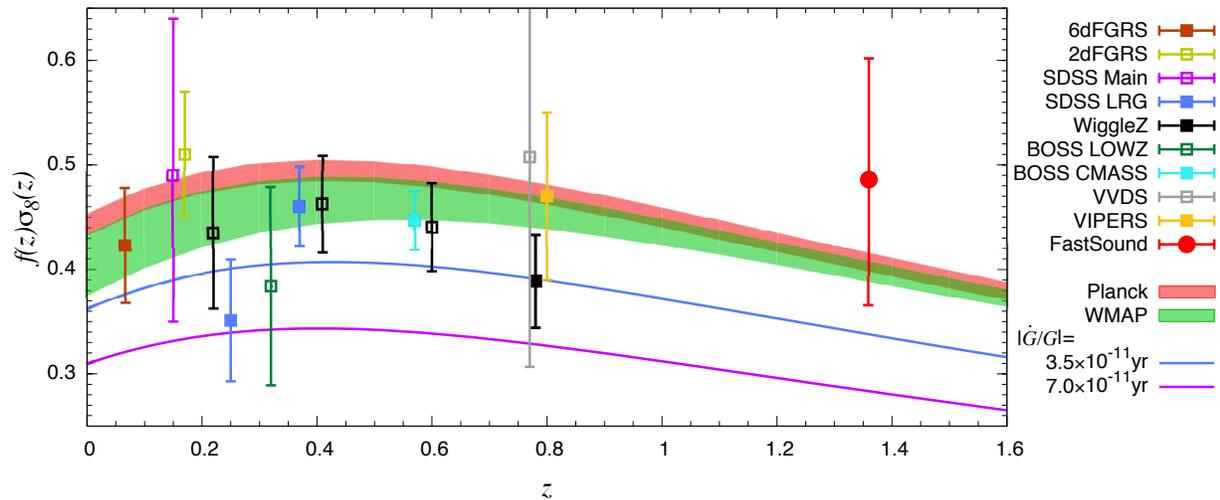}
  \end{center}
 \caption{Constraints on the growth rate $f\sigma_8$ as a function of redshift compared to the $\Lambda$CDM model with the best fitting models from the CMB experiments.
The data points are the same as those in figure \ref{fig:fsigma8_table_lss}. 
Theoretical predictions with 68\% confidence intervals based on WMAP9 and Planck CMB measurements 
are shown as the green and red shaded regions, respectively.
The early, time varying gravitational constant models with $\dot{G}/G=3.5\times 10^{-11} [{\rm year}^{-1}]$ and $7.0\times 10^{-11}[{\rm year}^{-1}]$ are respectively shown as the blue and magenta lines.  
}
\label{fig:fsigma8_table_cmb}
\end{figure*}

%
\subsubsection{Galileons}

Another class of models that modify gravity are based around a
scalar field, $\phi$ that satisfies a Galilean shift symmetry:
$\partial_{\mu} \phi \to \partial_{\mu} \phi+b_{\mu}$ 
in Minkowski space-time.
One can obtain general Lagrangians of  ``Galileons'' 
\citep{Nicolis:2009} and, in particular, do so in 
curved space-time leading to ``covariant Galileons'' \citep{Deffayet:2009}. 
The analytic estimation of $G_{\rm eff}$ and 
the full numerical integration of cosmological perturbations
for the covariant Galileon were first carried out
by \citet{de-Felice:2011a}. 
In the massless limit ($m_{\phi} \to 0$), the effective 
gravitational coupling can be schematically expressed in the 
form \citep{Tsujikawa:2015,Perenon:2015}
\begin{equation}
G_{\rm eff}=\frac{c_{\rm t}^2}{16\pi q_{\rm t}} 
\left( 1+ Q_{\rm s} \right)\,,
\label{Geffex}
\end{equation}
where $Q_{\rm s}$ describes the scalar-matter interaction, 
 $q_{\rm t}$ is associated with the no-ghost condition
of tensor perturbations \citep{Kobayashi:2011}, 
which is required to be positive and 
the quantity $c_{\rm t}^2$ corresponds to the tensor propagation 
speed squared, which needs to be positive to avoid the Laplacian 
instability. Using conditions for avoiding ghosts and Laplacian 
instabilities of both scalar and tensor perturbations, 
it follows that $Q_{\rm s} \geq 0$ \citep{Tsujikawa:2015}.

In figure \ref{fig:fsigma8_table_lss} we plot $f(z)\sigma_8(z)$ with the best-fit amplitude constrained by the RSD data
for the covariant Galileon as the dotted dashed line.
Here, we chose the parameters $(\alpha,\beta)= (1.347, 0.442)$ of the model
presented in \citet{Okada:2013} as an example;
these parameter values satisfy the theoretically consistent conditions
(such as the absence of ghosts and Laplacian instabilities).
The curve of $f(z)\sigma_8(z)$ constrained by the RSD data alone 
(without using the CMB constraint on $\sigma_8(0)$) exhibits significant 
difference from those of the $\Lambda$CDM and $f(R)$ gravity 
at high redshifts ($z>1$). Thus, the FastSound data is very useful to 
distinguish the covariant Galileon from other modified gravity  theories. 
One can generalize the covariant Galileon to the extended 
Galileon \citep{de-Felice:2012}.
In this case the growth rate of $\delta_m$ is typically 
greater than that in the $\Lambda$CDM model, but it is not as large as
that of the covariant Galileon \citep{Okada:2013}. 
In Fig.~\ref{fig:fsigma8_table_lss} we 
show the best-fit curve of $f(z)\sigma_8(z)$ for the extended 
Galileon as a dot-dashed line. Here we adopted the parameters $(\alpha,\beta)=(3.0,1.434)$ (see \cite{Okada:2013}).
The difference from the $\Lambda$CDM model is not so large, but 
it will be possible to discriminate between the two models in future 
high-precision observations.

\subsubsection{Early, time varying gravitational constant model}
\label{varyingmodel}

Finally, we step back from considering specific models that arise from 
fundamental Lagrangians; we now take our equation (\ref{f_eq}) and 
assume a simple functional for  $\xi$ such that it changed at early times 
(for example at $z\sim 10^2$) and then remained 
constant \citep{Baker:2014,Leonard:2015}. This leads
to a slow and steady modification of the growth rate all the way up-to $\Lambda$ domination. The cumulative will change the overall amplitude of 
$f\sigma_8$ for $z< 2$ such that, depending on how we normalize the overall amplitude of fluctuations, we will either be completely consistent or inconsistent with
RSD measurements. Normalizing to the the RSDs render such a theory effectively indistinguishable from $\Lambda$CDM  as we can see in  
figure \ref{fig:fsigma8_table_lss} where we show such a theory 
with a red-colored line  with $\dot{G}/G=3.5\times 10^{-11}[{\rm year}^{-1}]$.
This degeneracy comes from the fact that we obtained the best-fit 
curve of $f(z)\sigma_8(z)$ without employing the CMB bound 
of $\sigma_8(0)$.

\subsection{Consistency with CMB experiments}\label{sec:fsigma8_cmb}

Finally, we test the consistency of our bounds on $f(z)\sigma_8(z)$ from 
the FastSound survey with CMB anisotropy probes of $\sigma_8(0)$.
In figure \ref{fig:fsigma8_table_cmb}, together with the observational constraints the same as figure \ref{fig:fsigma8_table_lss}, we show the $1-\sigma$ confidence regions $f(z)\sigma_8(z)$ from the WMAP 9-yr data  \citep{Hinshaw:2013} and from the Planck data \citep{Planck-Collaboration:2015} using the green and red shaded bands. 
The prediction for the WMAP is the same as in figure 12 of \citet{More:2015}, while 
that for the Planck has been updated for the 2015 data.
Almost all the low-$z$ constraints of $f(z)\sigma_8(z)$ are in agreement with the two theoretical predictions for the $\Lambda$CDM model within $1-\sigma$ error, and all the rest are also consistent within $2-\sigma$. 
We find that the growth of density perturbations in the 
$\Lambda$CDM model, combined with the CMB bounds of $\sigma_8(0)$, is consistent with our result from the RSD of the FastSound survey as well as the low-$z$ probes at $z<1$. 

In figure \ref{fig:fsigma8_table_cmb}, we also plot $f(z)\sigma_8(z)$ in
the early, time varying gravitational constant models 
with $\dot{G}/G=3.5\times 10^{-11}[{\rm year}^{-1}]$ and 
$7.0\times 10^{-11}[{\rm year}^{-1}]$ for the normalization of 
$\sigma_8(0)$ consistent with CMB constraints 
(although they might lead to a large ISW) as the blue and 
magenta lines, respectively
As we have seen in section \ref{sec:fsigma8_mg}, the $f(z)\sigma_8(z)$ 
constraints alone cannot distinguish the varying $G$ model from 
the $\Lambda$CDM model, reflecting the fact that the ratio between
$f(z)\sigma_8(z)$ in these two models is nearly constant.
If we take into account the CMB normalization of $\sigma_8(0)$, 
the degeneracy of $f(z)\sigma_8(z)$ is broken. 
For larger $\dot{G}/G$ the values of $f(z)\sigma_8(z)$ tend to be smaller, 
so such cases can be severely constrained from the RSD data of 
the FastSound survey.

\section{Conclusion}\label{sec:conclusion}

The FastSound survey is a near-infrared galaxy redshift survey which probes the 3-d galaxy distribution at $1.19<z<1.55$ through H$\alpha$ emission lines obtained with FMOS spectrograph at Subaru Telescope.
In this paper we have analyzed the redshift-space clustering of galaxies, and found the coherent squashing effect in the anisotropic correlation function $\xi(r_p,r_\pi)$ known as the Kaiser effect and non-zero quadrupole moment $\xi_2(r)$ up to $r\sim 40\himpc$.
For the error estimation, we constructed the covariance matrix by generating 640 mock samples with the same geometry as the FastSound survey from $N$-body simulations. 

Limiting our analysis only at large scales, we adopted a simple model of the redshift-space correlation function, which is based on the linear Kaiser RSD factor times the nonlinear matter power spectrum with two free parameters, the growth rate parameter $f(z)\sigma_8(z)$ and the bias parameter $b(z)\sigma_8(z)$ at $z\sim 1.36$.
We obtained the constraint $f\sigma_8=0.482^{+0.116}_{-0.116}$ ($4.2\sigma$ CL) when the monopole and quadrupole moments at $8<r<80\himpc$ were used and $b\sigma_8$ is marginalized over. 
The constraint is consistent with the value predicted by the $\Lambda$CDM models with Einstein's general relativity, obtained by Planck (WMAP) $f\sigma_8\sim 0.392$. This is the first test of the gravity theories at the redshift range $1<z<2$, and also the first cosmological analysis using the Subaru Telescope.
The bias parameter of the FastSound galaxy sample is determined to $b(z)\sigma_8(z)=0.814^{+0.076}_{-0.080}$, corresponding to $b\sim 1.9$, consistent with the small-scale analysis based on a halo occupation modeling. 

Density perturbations grow differently with time under different gravity theories. 
We have thus demonstrated that our measurement of the growth rate at $z\sim 1.4$ is useful to distinguish modified gravity theories, independently of the CMB experiments, by combining with the low-$z$ measurements of the growth. 
Several ongoing and future galaxy redshift surveys which target the same redshift range, such as the extended BOSS (eBOSS) survey \citep{Dawson:2016}, and the Hobby-Eberly Telescope Dark Energy Experiment (HETDEX, \cite{Adams:2011}). There is also a project which targets the similar but a broader redshift range than the FastSound survey and uses the same telescope: The Prime Focus Spectrograph (PFS) of the Subaru Measurement of Images and Redshifts (SuMIRe) project \citep{Takada:2014}. These surveys will enable to test modified gravity models at the redshift $z>1$ with higher precision by RSD. The main sample targets of these surveys are emission line galaxies,
thus the constraint and analysis we provide in this paper will be useful and a basis of such larger surveys. Particularly, the techniques of estimating systematic effects such as the OH masks we have developed to construct the selection functions in this paper will be applicable and useful to the analysis of these survey data. 

The current analysis could be improved by considering following issues. 
First, in this analysis we considered only the clustering at large scales $r>8\himpc$ and used a simple model for the correlation function, which is the linear Kaiser factor multiplied by the nonlinear density power spectrum of matter in real space. By doing a more aggressive analysis with smaller $r_{\rm min}$ we could obtain a tighter constraint on $f\sigma_8$ while we need to treat the systematic effects of the data more carefully \hikagep. Moreover, we need to adopt more sophisticated theoretical models for the analysis for the nonlinearities to use smaller scale data. The current analysis has been performed assuming $\Lambda$CDM and GR, thus it is so-called a consistency test. We need to perform a test on modified gravity theories consistently using the theoretical model based on the same modified gravity (e.g., \cite{Taruya:2014,Song:2015}).
Second, the growth rate parameter $f\sigma_8$, that is sensitive to modified gravity theories, is strongly degenerate with the bias parameter $b\sigma_8$ as we have explain in section \ref{sec:basics} and have actually seen in figure \ref{fig:fsig8_bsig8}. The measurement of the higher-order statistics, e.g., the three-point correlation function, allows one to constrain the bias parameter independently of the two-point analysis presented in this paper. Combining the three-point and two-point correlation functions may tighten the constraint on the growth rate. Such analyses will be presented in future papers. 

\bigskip
We thank David Alonso for comments on the early version of this paper and Danielle Leonard for supplying us with some of the theoretical curves in figures \ref{fig:fsigma8_table_lss} and  \ref{fig:fsigma8_table_cmb}.
We also thank the anonymous referee for helpful comments.
The FastSound project was supported in part by MEXT/JSPS KAKENHI Grant Numbers 19740099, 19035005, 20040005, 22012005, and 23684007. 
TO acknowledges the support by Grant-in-Aid for Young Scientists (Start-up) from the Japan Society for the Promotion of Science (JSPS) (No. 26887012).
CH acknowledges the support by JSPS Grant-in-Aid for Young Scientist (B) Grant Number 24740160.
KG acknowledges support for this work from ARC Linkage International Fellowship LX0881951.
CB acknowledges the support of the Australian Research Council through the award of a Future Fellowship.
ST is supported by the Grant-in-Aid for Scientific Research Fund of the JSPS No. 24540286 and MEXT KAKENHI Grant-in-Aid for Scientific Research on Innovative Areas ``Cosmic Acceleration'' (No. 15H05890).

\appendix


\section{Systematic effects on determination of the growth rate}\label{sec:systematics}

In this appendix 
we examine how the treatments of various observational systematics alter the final constraints on $f\sigma_8$ and $b\sigma_8$. 
Here we use the data at separation $8<r<80\himpc$ and show the $1\sigma$ error bounds of the two parameters.
The top row of figure \ref{fig:fsig_systematic} labeled as [0] shows our fiducial constraints, the same as the results presented in section \ref{sec:analysis}, and  
the best-fitting values of $f\sigma_8=0.482$ and $b\sigma_8=0.814$ are shown as the vertical dotted lines as references for the following tests.

\subsection{Selection functions}\label{sec:sys_selection}
Let us start by looking at systematics due to methods of constructing the selection functions used in our analysis.
In the fiducial result [0], we constructed the radial selection function by fitting equation (\ref{eq:radial_3para}) to the binned redshift distribution of galaxies independently to each of the four CFHTLS survey fields (the solid lines in figure \ref{fig:zdist_w1-4}). 
As we have seen in section \ref{sec:data}, however, the radial distributions for the four fields are very similar with each other but not the same. 
We modify this method and fit the radial distribution of the galaxies summed over all the four fields (the solid line in figure \ref{fig:zdist}), and we show the resulting constraints in figure \ref{fig:fsig_systematic} denoted as [1]. 
Our fiducial results for ($f\sigma_8$, $b\sigma_8$) are shifted to slightly larger values, but the changes are negligibly small compared to the error bars.  

\begin{figure}\begin{center}\FigureFile(80mm,80mm){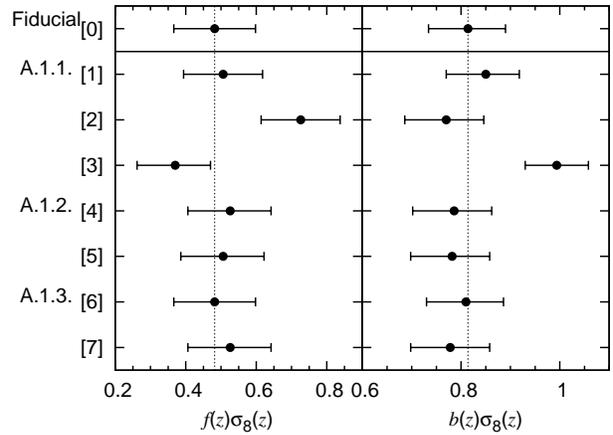}
  \end{center}
 \caption{Constraints on the growth rate $f\sigma_8$ ({\it left}) and the bias parameter $b\sigma_8$ ({\it right}) obtained by analyzing the data at $8<r<80\himpc$ 
 with various conditions. Error bars show the $1\sigma$ confidence levels. 
The first row labeled as [0] is our fiducial results presented in section \ref{sec:analysis}, and the rows [1]-[7] are the results where each of the conditions for the analysis is modified from the result [0]. 
The vertical dotted lines are the best fitting values for the fiducial results with $r_{\rm min}=6\himpc$. 
The row [1] is the case where we combined the radial distributions of galaxies for the four fields to construct the radial selection function. 
The result in the row [2] is obtained by neglecting the decrease of the detection efficiency of lines due to the OH masks.
The row [3] shows the result when the angular weight is assumed to be constant, namely the angular selection function is ignored. 
The row [4] is obtained when we apply the CFHT mask for the data only in $z$ band, while the row [5] is obtained without applying any of CFHT masks.
We used different parameter set for the fiber allocation correction, $(\theta_0,a)=(0.282,1.5)$ for the result [6], while the effect for the fiber allocation correction is ignored for the result [7]. 
}
\label{fig:fsig_systematic}
\end{figure}

The loss of the efficiency for detecting emission lines due to the OH masks was modeled using the observed deficit of galaxies near the masks [equation (\ref{eq:efficiency_loss})].
If the loss were ignored, we would measure artificial anisotropies at the wavelengths of the masks. 
In practice it produces the larger quadrupoles, which leads to $f\sigma_8$ biased to a larger value, as shown as the result [2] in figure \ref{fig:fsig_systematic}. It thus implies that estimating the decrease of the detection efficiency is essentially important to obtain an unbiased constraint on $f\sigma_8$. 

As detailed in section \ref{sec:angular}, the angular selection function for a FOV A, $W_A$, is estimated by the ratio of the number of galaxies with detected lines to that of all the targets. In the regions where two FOV's are overlapped, the angular selection function was estimated by equation (\ref{eq:angular_overlap}). 
The value of $W_A$ varies over different FOV's.
To see the importance of the estimation of $W_A$, we perform the analysis by assuming a constant weight $W_i={\rm constant}$. 
As expected, the resulting constraints are biased, particularly the anisotropic feature of the clustering is smeared, thus 
the constraint on $f\sigma_8$ is shifted to a smaller value and correspondingly that on $b\sigma_8$ is to a larger value, as shown in [3] in figure \ref{fig:fsig_systematic}.

In constructing the selection functions, we assumed a universal radial
selection function; namely the radial and angular selection functions
are not correlated. As shown in figure 5 of Paper II, however, there
is a large scatter of the observed line flux for the line $S/N$ above
the threshold of 4.5 adopted in our analysis, and this would be
related to the wide variation of observing conditions.  To test the
assumption of the universal radial selection function, we divide our
galaxy sample into the two subsamples separated by the median (0.08)
of $W=N^{\rm det}/N^{\rm tar}$, which is a good indicator of observing
condition and used in estimating angular selection funciton.  In
figure \ref{fig:zdist_angular}, we show the radial distributions of
galaxies for these subsamples.  The two distributions are very
similar, though the large scatter, and they are consistent within
$1-\sigma$ except for a few bins. The difference of the mean redshift
for the two subsamples is less than $1\%$, and much smaller than the
differences among different fields (fig \ref{fig:zdist_w1-4}). It is
somewhat expected because at any redshift the majority of the galaxies
should have the line flux close to the detection limit. We thus
conclude that our result is not very sensitive to the observing
conditions, and adopting the angularly-independent radial selection
function is a reasonable approximation.

\begin{figure}\begin{center}\FigureFile(70mm,70mm){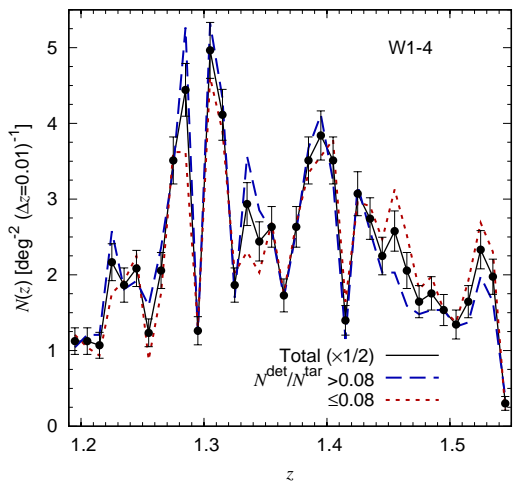}
\end{center}
 \caption{Redshift distributions of the two galaxy subsamples split by
   the median angular selection function, $W=N^{\rm det}/N^{\rm tar}=0.08$.
   As in figure \ref{fig:zdist}, signal-to-noise ratio threshold of
   $S/N>4.5$ is adopted for emission lines.  The black line with the
   Poisson error is the distribution for the whole sample (the same as
   that in figure \ref{fig:zdist}, but divided by two to compare with
   the subsamples).  The difference of the mean redshifts for the two
   subsamples is less than 1\%.  }
\label{fig:zdist_angular}\end{figure}

\subsection{Data at CFHTLS mask regions}\label{sec:sys_cfhtls}

For our main analysis [0] we applied the whole 5 bands of the CFHTLS mask, 
and excluded all the galaxies masked by any of the 5 bands.
Here we perform the analysis after excluding only the galaxies masked in $z$-band, 
and the result is shown in figure \ref{fig:fsig_systematic} and labeled as [4]. 
The result without applying any CFHTLS mask is also shown as the label [5]. 
As seen in the figure, the cosmological constraints are not significantly biased for the two cases.
Nevertheless, to be conservative we decided to remove all the galaxies masked in any of the 5 bands in our 
main analysis because all the 5 bands are used to estimate photometric redshifts and 
H$\alpha$ fluxes in the target selection processes, as mentioned in section \ref{sec:fastsound}.

\subsection{Fiber allocation correction}\label{sec:sys_fiber_alloc}
In measuring the correlation function of the FastSound galaxies we took into account the effect of fiber allocation failures in 
section \ref{sec:fiber_alloc}. We used the simple model [equation (\ref{eq:fiber_alloc})] with parameters fitted for $0.3<\theta < 50$ [arcmin]. 
To see the sensitivity of the fitting range to the cosmological result, we perform the same analysis by changing the minimum angular scale to 
0.7 [arcmin]. The best fitting parameters are $(\theta_0,a)=(0.282,1.5)$ and the function is shown as the dashed line in figure \ref{fig:fiberalloc}.
The constraints on $f\sigma_8$ and $b\sigma_8$ are shown as the label [6] in figure \ref{fig:fsig_systematic}. 
Although the shape of the fitting function changes, the constraints are almost unchanged, thus the fitting range does not significantly bias the result. 
Finally the result [7] is obtained when the fiber allocation failures are not corrected for, which corresponds to set equation (\ref{eq:fiber_alloc}) to unity. 
The result indicates that ignoring this effect would bias the cosmological fits.

\bibliography{ms.bbl}
\bibliographystyle{apj}

\end{document}